\documentclass[11pt,a4paper]{article}
\usepackage[utf8x]{inputenc}
\usepackage[T1]{fontenc}
\usepackage{mathptmx} 

\usepackage[margin=25truemm]{geometry}
\usepackage[pdftex]{graphicx} 
\usepackage{calc} 
\usepackage{authblk}
\usepackage{bm}
\usepackage{enumitem} 
\usepackage{braket,mathtools,amsmath,amssymb,subcaption}
\usepackage{authblk}
\usepackage[all]{nowidow} 
\usepackage{url}

\usepackage{lipsum} 

\usepackage{framed} 
\usepackage[framed]{ntheorem}
\newframedtheorem{frm-thm}{Theorem}
\usepackage{hyperref}
\hypersetup{colorlinks,linkcolor=blue}

\def\hx{\hat{x}}
\def\hy{\hat{y}}
\def\hz{\hat{z}}

\def\br{\mathbf{r}}
\def\bp{\mathbf{p}}
\def\fr{\frac{1}{2}}

\def\jx{\mathbf{j}_x}
\def\jy{\mathbf{j}_y}
\def\jz{\mathbf{j}_z}
\begin{document} 
\title{Multipole and fracton topological order\\ via gauging foliated SPT phases}

\author{Hiromi Ebisu$^1$}
\author{Masazumi Honda$^2$}
\author{Taiichi Nakanishi$^{2,1}$}
\affil{$^1$Yukawa Institute for Theoretical Physics, Kyoto University}
\affil{$^2$Interdisciplinary Theoretical and Mathematical Sciences Program (iTHEMS)}
\maketitle
\thispagestyle{empty}

\begin{abstract}
Spurred by recent development of fracton topological phases, unusual topological phases
possessing fractionalized quasi-particles with mobility constraints, the concept of symmetries has been renewed. In particular, in accordance with the progress of multipole symmetries, associated with
conservation of multipoles, such as dipole or quadruple moments as well as global charges,
there have been proposed topological phases with such symmetries. These topological phases are
unconventional as excitations are subject to mobility constraints corresponding to the multipole
symmetries. We demonstrate a way to construct such phases by preparing layers of symmetry protected topological (SPT) phases and implementing gauging a global symmetry. After gauging, the
statistics of a fractional excitation is altered when crossing the SPT phases, resulting in
topological phases with the
multipole symmetries. The way we construct the phases allows us to have a comprehensive understanding of field theories of topological phases with the multipole symmetries and other fracton models.

\end{abstract}

\newpage
\pagenumbering{arabic}
\setcounter{page}{1}
\tableofcontents

\section{Introduction}
Topologically ordered phases have been one of the central subjects in condensed matter physics, and provide us a paradigm shift in understating phases of matter~\cite{Tsui:1982yy,Kalmeyer1987,wen1989chiral,read1991large}. 
The most prominent feature of these phases is that they support fractionalized quasi-particle excitations, the so-called anyons~\cite{leinaas1977theory,Laughlin1983}.  
Such exotic excitations are not only theoretically and experimentally intriguing but also are of practically importance for utilizing quantum computers~\cite{KITAEV20032,dennis2002topological,Nayak2008, Mong:2013zda}. \par
Recently, there have been proposed new topological phases, referred to as the fracton topological phases~\cite{chamon,Haah2011,Vijay}. 
The distinct property compared with the conventional topologically ordered phases is that the fractional excitations are subject to mobility constraints. 
Due to this feature, ground state degeneracy (GSD) becomes sub-extensive. 
Compared with the conventional topologically ordered phases, where the GSD depends on topology of a manifold, the sub-extensive GSD found in the fracton topological phases implies that one cannot separate the UV and IR physics, which necessitates a new theoretical framework. \par
To aim for establishing a consistent continuum field theoretical description of the fracton topological phases, recently the concept of symmetries has been reconsidered. 
One of such update is \textit{multipole symmetry}, which is a generalization of ordinary global symmetry in the sense that multipole moments, such as dipole or quadrupole moments in addition to global charges are conserved~\cite{griffin2015scalar,Pretko:2018jbi,PhysRevX.9.031035,PhysRevB.98.035111,PhysRevB.106.045112,Jain:2021ibh}. 
More explicitly, multipole symmetries correspond to a case when a theory is invariant under a global phase rotation which has a polynomial form of the spatial coordinate.
For instance, in the case of a scalar theory with a field~$\Phi$, we say the theory respects dipole symmetry when the theory is invariant under global phase shift in the form $\Phi\to e^{i(\alpha+x\cdot\beta)}\Phi$, where $\alpha$, $\beta$, and $x$ represent a constant number, a constant spatial vector, and the linear spatial coordinate vector, respectively. 
Other cases of higher order multipole symmetries, such as quadrupole symmetry can be analogously defined, by considering the phase shift which depends on higher order spatial coordinates. Motivated by this progress,  
several topological phases with such a symmetry were studied~\cite{PhysRevB.106.045112,PhysRevB.106.045145,oh2022rank,ebisu2209anisotropic,PhysRevB.107.125154}. These phases have fractional excitations which are subject to a mobility constraint, corresponding to the multipole symmetry. Due to this feature, the phases exhibit unusual GSD dependence on the system size; the GSD depends on the greatest common divisor between integer $N$ characterizing the fractional charge and the system size.
Although several gauge theories and topological models with such symmetries have been recently studied,
it is still largely unknown that what physical implication or what kind of physical properties one can learn when a theory respects such symmetries. \par
In this paper, we demonstrate a way to establish topological phases with the multipole symmetries. 
In doing so, we prepare stack of symmetry protected topological (SPT) phases~\cite{doi:10.1126/science.1227224,spt2013,chen2014symmetry} with global~$\mathbb{Z}_N$ symmetry, which are invertible gapped phases respecting the global symmetry~$\mathbb{Z}_N$, and implement gauging.
Generically, it is known that via gauging, a $d+1$-dimensional SPT phase with global symmetry $G$, characterized by a $(d+1)$-cohomology class 
is mapped to the so-called Dijkgraaf-Witten theory~\cite{DW1990} with a twist term, corresponding to the non-trivial cohomology class~\cite{propitius1995topological,PhysRevB.86.115109,PhysRevLett.114.031601,PhysRevB.91.165119}. 
Such a twist term, that we dub the \textit{Dijkgraaf-Witten (DW) twist} term throughout this paper, modifies braiding statistics of anyons in the theory. 
One example of the DW twist terms defined in $(2+1)$d\footnote{
Throughout this paper, we use two letters ``d" and ``D" to abbreviate the dimensions. The small ``d" stands for space-time dimensions whereas the capital one ``D" does for spatial dimensions.
}  has the form $\sim a^1\wedge a^2\wedge a^3$ where $a^j\;(j=1,2,3)$ denotes a $1$-form gauge field associated with a global symmetry~\cite{propitius1995topological,PhysRevB.86.115109,PhysRevLett.114.031601,PhysRevB.91.165119}.
Key insight of our construction is that via gauging the global symmetry, the stack of the SPT phases become arrays of the DW twist terms 
to change statistics of quasi-particles, which results in a topological phase with the multipole symmetries. 
By this approach, one can make intuitive understanding of the BF theories with multipole symmetries that was previously discussed~\cite{2023arXiv231006701E}. 
To our knowledge, the present paper is the first work to address how the DW twist term plays a pivotal role in the context of the fracton topological phases, especially, 
topological phases with multipole symmetries.
\par
Our proposal may share the same spirit as an attempt to construct fracton topological models, such as the X-cube model via defect condensations~\cite{PhysRevResearch.2.043165}. 
Compared with this attempt, our construction is simpler as we do not impose energetic projection on the models to implement the condensation. 
Our approach allows us to make comprehensive understanding of foliated BF theories~\cite{foliated1,foliated2,Ohmori:2022rzz,Spieler:2023wkz,Cao:2023rrb,2023arXiv231006701E}, which are special topological field theories describing fracton and multipole topological order by interpreting coupling terms $A^I\wedge b\wedge e^I$ (to be defined in the next section) as the foliated DW twist terms. 
Moreover, our result would comply with growing interests in 
unified construction of topological defect lines obtained by stacking SPT phases and gauging~\cite{10.21468/SciPostPhys.15.3.122}, implying a possibility to explore the multipole topological phases in a broader context. 
As a byproduct, our construction of the stacking SPT phases and gauging leads us to the other fracton model, such as the exotic $\mathbb{Z}_N$ gauge theory~\cite{seiberg2021exotic} and X-cube model~\cite{Vijay}. 
In this regard, the way we construct phases may provide us a unified insight on topological phases with multipole symmetries and other fracton models.
\par
The rest of this work is organized as follows. In Sec.~\ref{s2}, to extract an intuition behind our construction, we introduce a BF theory description of topological phases with multipole symmetries, following the procedure discussed in~\cite{2023arXiv231006701E}. In Sec.~\ref{s3}, we study a UV lattice model to construct the topological phase with the multipole symmetry, especially dipole symmetry in the $x$-direction, which is the simplest case of the multipole symmetry. We explicitly demonstrate how the DW twist term affects the statistics of fractional excitations, giving the desired topological phase with the dipole symmetry. We also generalize the discussion to the other case of the multipole symmetry in Sec.~\ref{s4}. In Sec.~\ref{s5}, we apply our approach to other fracton models, the so-called exotic $\mathbb{Z}_N$ gauge theory~\cite{seiberg2021exotic} and the $X$-cube model~\cite{Vijay}. Finally, we conclude our work in Sec.~\ref{s6} with a few remarks and future research directions.

\section{BF theory with dipole symmetry}
\label{s2}
To gain intuition behind our idea to construct topological phases with multipole symmetry via gauging SPT phases, we begin by reviewing how topological BF theories with multipole symmetries are introduced, starting with an argument of these symmetries and gauge fields associated with them. 
Throughout this section, we focus on $(2+1)$d system
and for the sake of the notational brevity, we employ the differential forms.\par
Suppose we have a theory with conserved charges associated with global $U(1)$ and dipole symmetry in the $x$-direction, described by $Q$ and $Q_{x}$ respectively. 
These charges are subject to the following relation~\cite{hirono2022symmetry,PhysRevB.106.045112,2023arXiv231006701E}
\begin{equation}
 [iP_I,Q]=0 \quad   [P_x,Q_x]=Q,\quad [P_y,Q_x]=0 \qquad (I=x,y).
 \label{eq:re2}
\end{equation}
Here, $P_I$ denotes the translation operator along the $I$-th direction.
While the first relation is obvious as~$Q$ represents the global charge, the intuition behind the second and third relations is that one associates the charge~$Q_x$ with the dipole moment in the $x$-direction, $x\rho$. 
Indeed, by writing the charge and dipole moment as $\rho$ and $x\rho$, where~$\rho$ denotes the density of the $U(1)$ charge, we think of shifting them by a constant in $x$- or $y$-direction. 
For instance, if one shifts the dipole moment $x\rho$ by a constant in the $x$-direction, the change of dipole moment under the shift is given by $(x+\Delta x)\rho-x\rho=\Delta x\rho$, where $\Delta x$ is constant, corresponding to the nontrivial commutation relation as described in the second relation in~\eqref{eq:re2}. 
The third relation is analogously discussed.\par
Introducing $1$-form currents $j$ and $K_x$,
we write the charges $Q$ and $Q_x$, via the integral expression using the conserved current as
\begin{equation*}
   Q=\int_V*j,\quad  Q_x=\int_V*K_x.
\end{equation*}
Here, $*$ represents the Hodge dual and $V$ does spatial volume.
To reproduce the relation~\eqref{eq:re2}, we demand that 
\begin{equation}
    *K_x=*k_x-x*j
\end{equation}
with
$k_x$ being a local (non-conserved) current. A simple calculation verifies the relation~\eqref{eq:re2}.
We also introduce $U(1)$ $1$-form gauge fields $a$, $A$ with the coupling term defined by
\begin{equation}
    S_{\rm dipole}=\int_V  \left(  a\wedge *j+A\wedge *k_x \right) .
\end{equation}
We need to have an appropriate gauge transformation such that the condition of the coupling term being
gauge invariant yields the conservation law of the currents.
The following gauge transformation does the job:
\begin{equation}
    a\to a+d\Lambda+\Sigma dx,\quad  A\to A+d\Sigma.\label{gaugetr1}
\end{equation}
Here, $\Lambda$ and $\Sigma$ denote the gauge transformation parameters.
One can check that the gauge invariance of the coupling term~$ S_{\rm dipole}$ under the gauge transformation~\eqref{gaugetr1} gives rise to
\begin{equation*}
   d*j=0,\quad  d\left( *k_x-x*j \right) =d*K_x=0.
\end{equation*}
\par
Now we introduce a BF theory using the gauge fields of the multipole symmetry. We define the following gauge invariant fluxes as 
\begin{equation}
    f\vcentcolon=da-A\wedge dx,\quad  F\vcentcolon=dA,
\end{equation}
and
put these fluxes in the BF theory format:
\begin{equation}
    \mathcal{L}_{\rm dipole}
    =\frac{N}{2\pi}b\wedge f+\frac{N}{2\pi}c\wedge F,\label{hh}
\end{equation}
with $b$ and $c$ representing $U(1)$ $1$-form gauge fields.
We also define the foliation field, which is introduced in the context of field theory of fracton topological phases, following~\cite{foliated1,foliated2}.
Generally, foliation is defined by co-dimension one sub-manifolds stacked along a direction, characterized by the $1$-form field, which we call the foliation field. 
In the present case, we set the foliation field by $e^x\vcentcolon=dx$.
Rewriting~\eqref{hh}, we arrive at the following BF theory:
\begin{equation}
   \boxed{\mathcal{L}_{\rm dipole}=\frac{N}{2\pi}a\wedge db+\frac{N}{2\pi}A\wedge dc+\frac{N}{2\pi}A\wedge b\wedge e^x.}\label{foliation dipole}
\end{equation}
The BF theory consists of two layers of the toric codes and the couplings between the two. Generally, the BF theory of a topological phase with multipole symmetries has the similar form as~\eqref{foliation dipole}; the BF theory consists of the layers of the toric codes with couplings between the layers, where the number of layers corresponds to the number of degrees of freedom of the multipoles (meaning, the number of global charge, dipole, quadrupole,~e.t.c.)~\cite{2023arXiv231006701E}.
%
In addition to the gauge symmetry~\eqref{gaugetr1}, the BF theory~\eqref{foliation dipole} also admits the following gauge symmetry with respect to $b$ and $c$:
\begin{equation}
    b\to b+d\lambda,\quad c\to c+d\gamma-\lambda e^x . \label{gaugetr3}
\end{equation}
with $\lambda$ and $\gamma$ being gauge parameters.
\par
To extract physical intuitions behind this BF theory, let us rewrite the theory in terms of tensor gauge fields.
One of the approaches for that is to integrate out the coupling term, corresponding to the last term in~\eqref{foliation dipole}, which put a several constraints on the gauge fields.
Integrating out $b_0$ gives the following condition:
\begin{equation}
  \partial_xa_y-\partial_ya_x=-A_y .
  \label{1}
\end{equation}
We also integrate out the 
$A_0$, and obtain the following condition
\begin{equation}
\partial_xc_y-\partial_yc_x=b_y.\label{2}
\end{equation}
One can eliminate the gauge fields $b_y$ and $A_y$ by
substituting the relations~\eqref{1}\eqref{2} into~\eqref{foliation dipole}.
Furthermore, we
introduce gauge fields by 
\begin{equation}
    A_{(xx)}\vcentcolon=\partial_xa_x-A_x,\quad
    c_{(xx)}\vcentcolon=\partial_xc_x+b_x ,
    \label{gauge3}
\end{equation}
whose gauge transformation reads [referring to~\eqref{gaugetr1}\eqref{gaugetr3}]
\begin{eqnarray}
    A_{(xx)}\to A_{(xx)}+\partial_x^2\Lambda,\quad
    c_{(xx)}\to c_{(xx)}+\partial_x^2\gamma .
\end{eqnarray}
After the substitution, the Lagrangian~\eqref{foliation dipole} has the following form
\begin{eqnarray}
    \mathcal{L}_{\rm dipole}
   &=&\frac{N}{2\pi}\biggl[a_0(\partial_x^2c_y-\partial_yc_{(xx)})+A_{(xx)}(\partial_tc_y-\partial_yc_0)
+
a_{y}(\partial_tc_{(xx)}-\partial_x^2c_0)\biggr].\label{Bfdi}
\end{eqnarray}
The BF theory~\eqref{Bfdi} is reminiscent of the one of the toric code~\cite{KITAEV20032} with a crucial difference being that the spatial derivative operator in the $x$-direction is replaced with the second order one.\par
 One of the challenges to study the BF theory such as the one in~\eqref{Bfdi} and those of other fracton models is the presence of the higher order spatial derivatives, yielding the UV/IR mixing. 
To circumvent this problem, we
follow an approach alluded in~\cite{Gorantla:2021svj,Gorantla:2021svj,PhysRevB.106.045112}. Instead of directly investigating the theory~\eqref{Bfdi}, 
we consider the following BF theory defined on a discrete Euclidean lattice where each gauge field takes integer value (see ~\cite{Gorantla:2021svj,Gorantla:2021svj,PhysRevB.106.045112,2023arXiv231006701E} for more detailed discussion on this point )
 \begin{eqnarray}
    \hat{\mathcal{L}}_{\rm dipole}
   &=&\frac{2\pi}{N}\biggl[\hat{a}_0(\Delta_x^2\hat{c}_y-\Delta_y\hat{c}_{(xx)})+\hat{A}_{(xx)}(\Delta_\tau\hat{c}_y-\Delta_y\hat{c}_0)
+\hat{a}_{y}(\Delta_\tau \hat{c}_{(xx)}-\Delta_x^2\hat{c}_0)\biggr].
\label{Bfdi0}
\end{eqnarray}
Here, the gauge fields take values in $\mathbb{Z}_N$ and $\Delta_{\tau,x,y}$ denotes the discretized differential operator.
This BF theory admits the following gauge symmetry:
\begin{eqnarray}
    \hat{a}_0\to\hat{a}_0+\Delta_\tau \lambda_a,\quad 
    \hat{A}_{(xx)}\to \hat{A}_{(xx)}+\Delta_x^2\lambda_a,\quad
    \hat{a}_{y}\to \hat{a}_{y}+\Delta_y\lambda_a\nonumber\\
    \hat{c}_0\to\hat{c}_0+\Delta_\tau \lambda_c,\quad 
    \hat{c}_{(xx)}\to \hat{c}_{(xx)}+\Delta_x^2\lambda_c,\quad
    \hat{c}_{y}\to \hat{c}_{y}+\Delta_y\lambda_c,
\end{eqnarray}
where $\lambda_a$ and $\lambda_c$ are integer gauge parameters.
Equation of motions of the theory~\eqref{Bfdi0} implies that the following gauge invariant field strength vanish
\begin{eqnarray}
    E_{ax}=\Delta_{\tau}\hat{A}_{(xx)}-\Delta_x^2\hat{a}_0,\quad
    E_{ay}=\Delta_{\tau}\hat{a}_{y}-\Delta_y\hat{a}_0,\quad
    B_a=\Delta_x^2\hat{a}_y-\Delta_y\hat{A}_{(xx)}, \nonumber\\
    E_{cx}=\Delta_{\tau}\hat{c}_{(xx)}-\Delta_x^2\hat{c}_0,\quad
    E_{cy}=\Delta_{\tau}\hat{c}_{y}-\Delta_y\hat{c}_0,\quad
    B_c=\Delta_x^2\hat{c}_y-\Delta_y\hat{c}_{(xx)}.
    \label{eom}
\end{eqnarray}
 The equation of motions ensures that there is no non-trivial local gauge invariant operators. 
Yet, the theory~\eqref{Bfdi0} does have
non-local gauge-invariant operators, which can be constructed from the gauge fields either ($\hat{a}_0,\hat{A}_{(xx)},\hat{a}_y$) or ($\hat{c}_0,\hat{c}_{(xx)},\hat{c}_y$). Especially, when the theory~\eqref{Bfdi0} is placed on a torus, it admits non-contractible Wilson loops, contributing to the GSD. 
To evaluate this number, we set the theory~\eqref{Bfdi0} on a torus geometry by imposing the periodic boundary condition via $\hat{x}\sim\hat{x}+L_x$ $\hat{y}\sim\hat{y}+L_y$,\footnote{
The coordinate $(\hx,\hy)$ takes integer number in the unit of the lattice spacing.
} i.e., the torus with the length in the $x/y$-direction being $L_x/L_y$. Focusing on the Wilson loops of the gauge fields $(\hat{A}_{(xx)},\hat{a}_y)$, the distinct number of such loops which amounts to be the GSD. In the $x$-direction, we have the following two types of the Wilson loops~\cite{ebisu2209anisotropic,2023arXiv231006701E}
\begin{equation}
    W_0(\hy)=\exp\biggl[\frac{2\pi i}{N}\sum_{\hx=1}^{L_x}\hat{A}_{(xx)}(\hx,\hy)\biggr],\quad
    W_{\rm dipole}(\hy)=\exp\biggl[\frac{2\pi i}{N}\alpha_x\sum_{\hx=1}^{L_x}\hx\hat{A}_{(xx)}(\hx,\hy)\biggr],\label{dp1}
\end{equation}
where $\alpha_x=\frac{N}{\gcd(N,L_x)}$ and $\gcd$ stands for the greatest common divisor. While the first term in~\eqref{dp1} corresponds to the standard Wilson loop which is also found in other conventional topologically ordered phases, the second term has the unconventional form. It is
interpreted as the ``dipole of the Wilson loop" in the sense that the argument of the exponential is a linear function of the coordinate $\hx$, reminiscent of the integration of the dipole moment, $x\rho$. This mirrors the fact that the BF theory respects the dipole symmetry in the $x$-direction~\eqref{eq:re2}. 
From equation of motions~\eqref{eom}, it follows that the two types of loops are deformable in the $y$-direction, implying that $ W_0(\hy)$ and $W_{\rm dipole}(\hy)$ do not depend on $\hy$. 
Hence, the non-contractible Wilson loops of the gauge field $\hat{A}_{(xx)}$ in the $x$-direction are labeled by $\mathbb{Z}_N\times\mathbb{Z}_{\gcd(N,L_x)}$. \par
One can evaluate the non-contractible Wilson loops in the $y$-direction, which are also labeled by the same quantum number, $\mathbb{Z}_N\times\mathbb{Z}_{\gcd(N,L_x)}$~\cite{ebisu2209anisotropic}. Overall, the GSD, which is equivalent to the number of distinct non-contractible Wilson loops, is given by
\begin{equation}
    {\rm GSD} =\left[ N\times\gcd(N,L_x) \right]^2.
    \label{gsd}
\end{equation}
Compared with the other fracton models, such as the $X$-cube model~\cite{Vijay}, which exhibit the sub-extensive GSD, the theory shows unusual GSD dependence on the system size due to the dipole symmetry.

\section{Gauging SPT phases}
\label{s3}
Now we come to the main part of this paper. 
In the previous section, based on the argument on the global and dipole symmetry, we introduced the topological field theory~\eqref{foliation dipole}, which respects such symmetries.
To further study the properties of the theory, we
integrate out the coupling term, $A\wedge b\wedge e^x$, corresponding to the third term in~\eqref{foliation dipole} to simplify the theory to the one with higher rank derivative terms~\eqref{Bfdi}.
Instead of integrating out the coupling term, in this section, we give more physical comprehension on this term and based on this understanding, we construct a UV lattice model corresponding to the topological field theory with the dipole symmetry.\par
The form $A\wedge b$ that enters~\eqref{foliation dipole} reminds us of the so-called the DW twist term~\cite{DW1990,propitius1995topological,PhysRevB.86.115109,PhysRevLett.114.031601,PhysRevB.91.165119}. 
Recalling that the $e^x$ is the foliated field, along which $(1+1)$d sub-manifolds labeled by $(t,y)$ are stacked,
it is tempting to interpret the BF theory~\eqref{foliation dipole} as the one described by the two copies of the standard BF theories with ``foliated DW twists", namely, DW twist terms $A\wedge b$ stacked in the $x$-direction. In the following, we demonstrate a concrete lattice model to realize such twist terms.

\subsection{Dipole symmetry in $x$-direction}
\label{2.2}
\begin{figure}[th]
    \begin{center}
      \begin{subfigure}[h]{0.34\textwidth}
  \includegraphics[width=\textwidth]{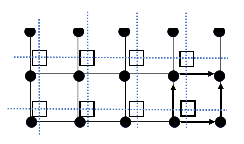}
         \caption{}\label{sfa}
             \end{subfigure}
        \begin{subfigure}[h]{0.26\textwidth}
    \includegraphics[width=\textwidth]{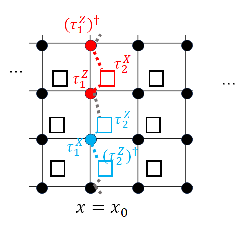}
         \caption{}\label{sfb}
             \end{subfigure} 
                   \begin{subfigure}[h]{0.58\textwidth}
    \includegraphics[width=\textwidth]{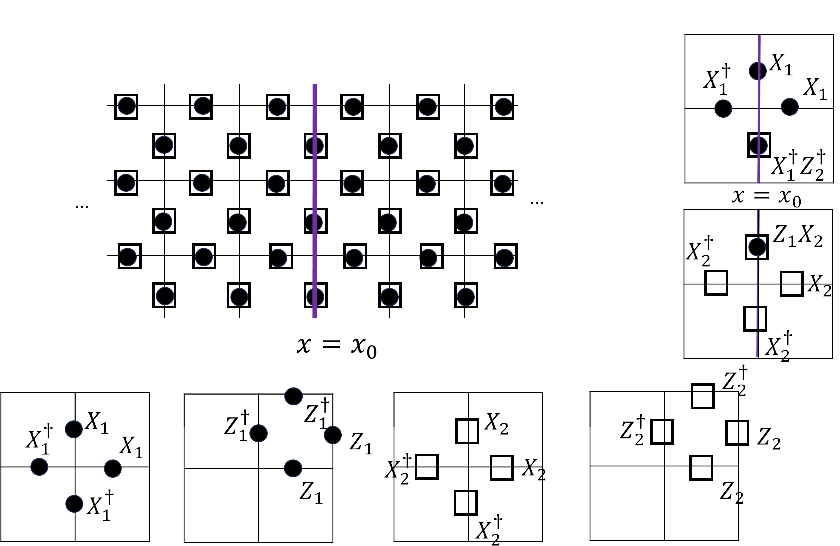}
         \caption{}\label{tc1}
             \end{subfigure} 
               \begin{subfigure}[h]{0.38\textwidth}
  \includegraphics[width=\textwidth]{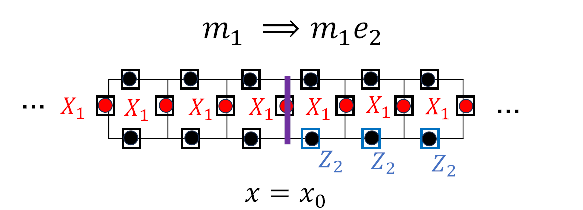}
         \caption{}\label{tj1}
             \end{subfigure}
\end{center}
\caption{(a) Two square lattices (solid black and dashed blue lines). The links are oriented according to the arrows. For illustrative purposes, we intentionally shift a bit one of the lattice with respect to the other.
(b) Along a line at $\hx=\hx_0$, we implement the CZ gates~\eqref{cz} on the $\mathbb{Z}_N$ qubit states according to~\eqref{spt22}. Examples of the terms in~\eqref{spt22} are shown by red and blue colors. 
(c) After gauging, the resulting Hamiltonian becomes two copies of the $\mathbb{Z}_N$ toric codes. Away from the line at $\hx=\hx_0$, the Hamiltonian composes of the vertex and the plaquette terms, shown in the four panels at the bottom, whereas at $\hx=\hx_0$, the additional Pauli operator is attached to the term depicted by the two panels on the right. 
(d) Trajectory of the magnetic charge $m_1$ when crossing the line at $\hx=\hx_0$ in the model~\eqref{spt3}. 
}
   \end{figure}

To this end, we introduce two copies of 2D square lattices.
We define integer coordinate of a node of one lattice (black dot in Fig.~\ref{sfa}) and the one of the other lattice (blue square in Fig.~\ref{sfa}) by $(\hx,\hy)$ in the unit of the lattice spacing. For illustration purposes, we intentionally shift a bit the one of the lattice with respect to the other one, as depicted in Fig.~\ref{sfa}. 
Also, we accommodate $\mathbb{Z}_N$ qubit at each site of the two square lattices, as portrayed in Fig.~\ref{sfa}.
We label the Hilbert space and the $\mathbb{Z}_N$ Pauli operator at each site as $\ket{a}_i$ 
($a\in \mathbb{Z}_N$) and $\tau^{X/Z}_i$, where the subscript $i=1,2$ distinguishes which square lattice the site belongs to. The Pauli operators act on the Hilbert spaces as
\begin{equation}
   \tau^X_i\ket{a}_i=\ket{a+1}_i,\quad \tau^Z_i\ket{a}_i=\omega^{a}\ket{a}_i\quad (\omega :=e^{2\pi i/N}).\label{zn}
\end{equation}

In this setting, we define the following Hamiltonian of a paramagnet:
   \begin{equation}
    H_0=-\sum_{\br} \Bigl[ \tau^X_{1,\br}+\tau^X_{2,\br}\Bigr] .
    \label{hami0}
\end{equation}
Here, we have defined vectors to abbreviate the coordinates by $\br\vcentcolon=(\hx,\hy)$. 
We introduce non-trivial SPT phases in this model. For this purpose, we start with introducing one $(1+1)d$ SPT along a $y$-direction at $\hat{x}=\hat{x}_0$. 
We define the generalized controlled-Z
gate $CZ$ and $CZ^\dagger$ acting on two $\mathbb{Z}_N$ qubit states. 
Its action on the states is given by
\begin{equation}
    CZ\ket{a}\ket{b}=\omega^{ab}\ket{a}\ket{b},\quad  
    CZ^\dagger \ket{a}\ket{b} =\omega^{-ab}\ket{a}\ket{b}\qquad (a,b\in\mathbb{Z}_N).
    \label{cz}
\end{equation}
At $\hat{x}=\hat{x}_0$, we act the CZ gate on two adjacent $\mathbb{Z}_N$ qubit states, composed of one state from a site on a square lattice and the one on the other lattice. 
More explicitly, we consider the following operator
\begin{equation}
    U_{CZ}\vcentcolon
    =\prod_{\substack{\br \\ \hx=\hx_0}} \Bigl[ CZ^\dagger_{(1,\br),(2,\br)}CZ_{(1,\br),(2,\br-{\jy})} \Bigr]
\end{equation}
with $\mathbf{j}_y$ being a unit vector in the $y$-direction, $\mathbf{j}_y\vcentcolon=(0,1)$~[for latter use, we also define $\jx\vcentcolon=(1,0)$],
and act it on the Hamiltonian~\eqref{hami0}. Pictorially, the CZ gates act on the states along the zigzag path portrayed in Fig.~\ref{sfb}. 
After the operation, Hamiltonian~\eqref{hami0} is transformed as
\begin{eqnarray}
 H_0^{\prime} 
 &:=& U_{CZ}^\dagger H_0U_{CZ} \nonumber \\ 
 &=& -\sum_{\substack{\br \\ \hx\neq \hx_0}} \Bigl[ \tau^X_{1,\br}+\tau^X_{2,\br}\Bigr]
 -\sum_{\substack{\hy \\ \hx=\hx_0}} \Bigl[ \tau^Z_{2,\br}\tau^X_{1,\br}(\tau^Z)^\dagger_{2,\br-\jy}+(\tau^Z)^\dagger_{1,\br+\jy}\tau^X_{2,\br}\tau^Z_{1,\br} \Bigr] .   
 \label{spt22}
\end{eqnarray}
Example of the last two terms, which are generalized cluster states~\cite{doi:10.1126/science.1227224,spt2013,chen2014symmetry}, are depicted in~Fig.~\ref{sfb}.
\par
The Hamiltonian~\eqref{spt22} respects a $(\mathbb{Z}_N )^2$ global symmetry generated by the operator
\begin{equation}
    \prod_{\br} \tau^X_{i,\br} \quad (i=1,2),
    \label{sym00}
\end{equation}
which allows us to gauge it.
Below, we follow the procedure given in~\cite{PhysRevB.86.115109} to implement gauging the global symmetry.\footnote{
See also~\cite{Vijay} for the similar discussion in the context of the fracton topological phases.
} 
 Via gauging,
one promotes the global symmetry~\eqref{sym00} to a local symmetry. To do so, we first introduce gauge degree of freedom 
on each link, described by the Hilbert space $\ket{\varphi}_{i,\br\pm\jx/2},\ket{\varphi}_{i,\br\pm\jy/2}$, ($i=1,2$, $\varphi\in\mathbb{Z}_N$), where $i$ distinguishes the link of the two lattices, with $\mathbb{Z}_N$ Pauli operators $X_{i,\br\pm\jx/2},Z_{i,\br\pm\jx/2}$, $X_{i,\br\pm\jy/2},Z_{i,\br\pm\jy/2}$ that act on the links,
satisfying the similar relation as~\eqref{zn}. 
The original $\mathbb{Z}_N$ spin d.o.f on each node can be thought of as a matter field while the newly introduced one located on each link can be regarded as $\mathbb{Z}_N$ gauge field.\footnote{
In the terminology of lattice gauge theory, $Z$ corresponds to link variable and $X$ does to its conjugate, which is a discrete analog of an exponentiation of electric field.
} 
We impose the condition that the physical Hilbert space has to have the trivial eigenstate of the following Gauss law operators:\footnote{
Whether we put the dagger $\dagger$ or not on the last four terms in~\eqref{gauss} depends on the orientation of the links in accordance with Fig.~\ref{sfa}, namely, we put (do not put) the dagger on a operator on a link which emanates from (terminates at) the node $\br$.}
\begin{eqnarray}
     G_{i,\br}=\tau^X_{i,\br}\times X^\dagger_{i,\br+\frac{\jx}{2}}X^\dagger_{i,\br+\frac{\jy}{2}}X_{i,\br-\frac{\jx}{2}}X_{i,\br-\frac{\jy}{2}}\quad (i=1,2),
     \label{gauss}
\end{eqnarray}
that is, physical states $\ket{\rm phys}$ are subject to $G_{i,\br}\ket{\rm phys}=\ket{\rm phys}$. The operator~\eqref{gauss} corresponds to the local $\mathbb{Z}_N$ spin flip symmetry. 
We also minimally couple the quadratic spin coupling terms in the original Hamiltonian~\eqref{spt22} to the $\mathbb{Z}_N$ gauge field as
\begin{eqnarray}
    \tau^Z_{2,\br}(\tau^Z)^\dagger_{2,\br-\jy}\to   \tau^Z_{2,\br}Z^\dagger_{2,\br-\frac{\jy}{2}}(\tau^Z)^\dagger_{2,\br-\jy},\quad
    (\tau^Z)^\dagger_{1,\br+\jy}\tau^Z_{1,\br} \to  (\tau^Z)^\dagger_{1,\br+\jy}Z_{1,\br+\frac{\jy}{2}}\tau^Z_{1,\br}
\end{eqnarray}
 so that these terms commute with the Gauss law~\eqref{gauss}.
 \par
In the restricted physical state, satisfying $G_{i,\br}=1$, the original Hamiltonian~\eqref{spt22} now becomes
\begin{eqnarray}
    \tilde{H}_0-=\sum_{\substack{\br \\ \hx\neq \hx_0}} \Bigl[ V_{1,\br}+V_{2,\br}\Bigr]
 -\sum_{\substack{\hy \\ \hx=\hx_0}} \Bigl[ V_{1,\br}\tau^Z_{2,\br}Z^\dagger_{2,\br-\frac{\jy}{2}}(\tau^Z)^\dagger_{2,\br-\jy} +V_{2,\br}(\tau^Z)^\dagger_{1,\br+\jy}Z_{1,\br+\frac{\jy}{2}}\tau^X_{2,\br}\tau^Z_{1,\br}\Bigr],  \label{gauge23}
\end{eqnarray}
where 
\begin{eqnarray}
    V_{i,\br}\vcentcolon= X_{i,\br+\frac{\jx}{2}}X_{i,\br+\frac{\jy}{2}}X^\dagger_{i,\br-\frac{\jx}{2}}X^\dagger_{i,\br-\frac{\jy}{2}}.\label{vv}
\end{eqnarray}
Since $\tau^Z_{i,\br}$ commutes with~\eqref{gauge23}, we set $\tau^Z_{i,\br} =1$, which gives rise to Hamiltonian described by only the gauge fields.  
We further add the following term to the Hamiltonian
 \begin{eqnarray}
    -\sum_{i,\bp} P_{i,\bp},\quad P_{i,\bp}\vcentcolon=Z_{i,\bp-\frac{\jy}{2}}Z_{i,\bp+\frac{\jx}{2}}Z^{\dagger}_{i,\bp+\frac{\jy}{2}}Z^\dagger_{i,\bp-\frac{\jx}{2}},
    \label{pla}
 \end{eqnarray}
 which ensures the fluxless condition to make the gauge theory dynamically trivial. Also, we have defined the coordinate of a plaquette by $\bp\vcentcolon=(\hx+\fr,\hy+\fr)$.
 Overall, the gauged Hamiltonian reads
\begin{eqnarray}
  H_{\rm gauged}=
 -\sum_{\substack{\mathbf{r} \\ \hx\neq \hx_0}}[ V_{1,\mathbf{r}}+ V_{2,\mathbf{r}}]
 -\sum_{\bp}[P_{1,\bp}+P_{2,\bp}]
 -\sum_{\substack{\mathbf{r} \\ \hx= \hx_0}}[   \widetilde{V}_{1,\mathbf{r}}+   \widetilde{V}_{2,\mathbf{r}}]
 +h.c.,
 \label{spt3}
\end{eqnarray}
where $V_{i,\br}$ and $P_{i,\bp}$ are given in~\eqref{vv}\eqref{pla} and
\begin{eqnarray}
     \widetilde{V}_{1,\br}\vcentcolon=Z^\dagger_{2,\br-\frac{\jy}{2}}X_{1,\br+\frac{\jx}{2}}X_{1,\br+\frac{\jy}{2}}X^{\dagger}_{1,\br-\frac{\jx}{2}}X^{\dagger}_{1,\br-\frac{\jy}{2}},\nonumber\\
      \widetilde{V}_{2,\br}\vcentcolon=Z_{1,\br+\frac{\jy}{2}}X_{2,\br+\frac{\jx}{2}}X_{2,\br+\frac{\jy}{2}}X^{\dagger}_{2,\br-\frac{\jx}{2}}X^{\dagger}_{2,\br-\frac{\jy}{2}}.\label{mod}
\end{eqnarray}
These terms are shown in Fig.~\ref{tc1}.
It is straightforward to check individual term that constitutes the Hamiltonian~\eqref{spt3} commutes with one another.
The gauged Hamiltonian~\eqref{spt3} describes two copies of the~$\mathbb{Z}_N$ toric code~\cite{KITAEV20032}, composed of the familiar form of the vertex and plaquette terms, $V_{i,\br},P_{i,\bp}\;(i=1,2)$ except along the 1D line at $\hx=\hx_0$, where the a single $Z_1$ or $Z_2$ operator is attached to the vertex term~\eqref{mod}. \par
To see the difference between the standard $\mathbb{Z}_N$ toric code and our gauged Hamiltonian, let us investigate the behavior of the excitations of the model~\eqref{spt3}. Since the individual term entering the Hamiltonian commutes with each other, the ground state~$\ket{\Omega}$ satisfies
\begin{eqnarray}
     V_{i,\br}\ket{\Omega}= \widetilde{V}_{i,\br}\ket{\Omega}=\ket{\Omega}\quad \forall\br,i ,\nonumber\\
     {P}_{i,\bp}\ket{\Omega}=\ket{\Omega}=\ket{\Omega}\quad \forall\bp,i .
     \label{hey1}
\end{eqnarray}
To discuss the behavior of quasi-particle excitations in the model, we denote a fractional excitation carrying an electric charge which violates the first condition in~\eqref{hey1} with eigenvalue $\omega$ as $e_{i,\br}$.
Likewise, we label a fractional excitation carrying an magnetic charge, violating the second condition in~\eqref{hey1} with eigenvalue being $\omega$ by $m_{i,\bp}$.
Further, for latter convenience, we put the symbol $\overline{(\cdots ) }$ on the top of a charge to denote its conjugate. 
For instance, we write an electric charge violating the first condition in~\eqref{hey1} with eigenvalue $\omega^{-1}$ as $\overline{e}_{1,\br}$.
\par
Away from $\hx=\hx_0$, we think of applying a single operator at a vertical link $X_{1,\br+\frac{\jy}{2}}$ ($\hx\ll \hx_0$) on the ground state, giving a pair of magnetic charges. Let us write the magnetic charge schematically as $m_1$, omitting the coordinate dependence for simplicity.
One can stretch the magnetic charge by successive action of the $X_1$ operators on the ground state in the horizontal direction. However, when crossing at $\hat{x}=\hat{x}_0$ the magnetic charge $m_1$ is dressed by a electric charge $e_2$ as $X_{1,(\hx_0,\hy)}$ does not commute with $\widetilde{V}_{2,\br}|_{\hx=\hx_0}$, see also Fig~\ref{tj1}. Analogous line of thought indicates that the magnetic charge $m_2$ induced by the operator $P_{2,\bp}$ is dressed with the electric charge $\overline{e}_1$ when traveling across the line at $\hx=\hx_0$. Also, it is straightforward to see that electric charges are intact by the terms at $\hx=\hx_0$. Therefore, we find that 
\begin{equation}
    m_1\Rightarrow m_1e_2,\quad m_2\Rightarrow m_2\overline{e}_1,\quad 
    e_1\Rightarrow e_1,\quad e_2\Rightarrow e_2.
    \label{fs}
\end{equation}
Here, $\Rightarrow$ represents the fractional excitations crossing the line at $\hx=\hx_0$. This
implies that the terms at $\hx=\hx_0$~\eqref{mod} modify the statistics of the magnetic charges (see also \cite{YOSHIDA2017387} for the relevant discussion in a different context.). More concise discussion of this point based on fusion rules of fractional charges will be presented in the next subsection. \par

\begin{figure}[t]
    \begin{center}
       \begin{subfigure}[h]{0.24\textwidth}
  \includegraphics[width=\textwidth]{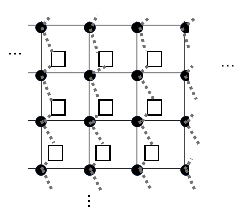}
         \caption{}\label{spt2}
             \end{subfigure}
                \begin{subfigure}[h]{0.34\textwidth}
  \includegraphics[width=\textwidth]{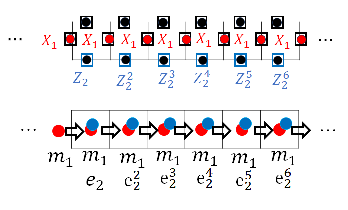}
         \caption{}\label{tj2}
             \end{subfigure}
        \end{center}
\caption{ (a) The same two lattices as Fig~\ref{sfa} with the CZ gates~\eqref{cz} being acted along the 1D zigzag grey dashed lines. (b) Trajectory of the magnetic excitation~$m_1$ dressed with the electric charges, $e_2$'s in the model described by the Hamiltonian~\eqref{dp}.}
   \end{figure}

After having seen the effect of the SPT phase, now we prepare arrays of such $(1+1)d$ SPT phases, which form stacked layers. We place the SPT phases so that each SPT goes along the $y$-direction and is located adjacent with one another in the $x$-direction. 
To this end, starting with the paramagnet Hamiltonian~\eqref{hami0}, we think of implementing the CZ gates defined in~\eqref{cz} on the states along the zigzag path demonstrated in Fig.~\ref{spt2},\footnote{
Similarly to Fig.~\ref{sfa}, in the actual configuration of Fig.~\ref{spt2}, nodes of the two lattices are located at the same position. However, we intentionally shift a bit one of the lattice with respect to the other for the sake of visual illustration.} and gauge the global $\mathbb{Z}_N$ symmetry.
A simple modification of the previous argument shows that the gauged Hamiltonian becomes
\begin{eqnarray}
      H_{\rm dipole} 
      =   
 -\sum_{\br}[    \widetilde{V}_{1,\br}+     \widetilde{V}_{2,\mathbf{r}}]
 -\sum_{\bp}[P_{1,\bp}+P_{2,\bp}]+h.c.,
  \label{dp}
\end{eqnarray}
where each term is given in~\eqref{pla} and \eqref{mod}.\par
Based on the discussion around~\eqref{fs}, one can see that every time the magnetic charge $m_1$ and $m_2$ go in the $x$-direction by one unit lattice spacing, they are dressed by an electric charge, giving
\begin{eqnarray}
     m_1 \xRightarrow[]{x} m_1e_2\xRightarrow[]{x} m_1e_2^2\xRightarrow[]{x} m_1e_2^3\xRightarrow[]{x}\cdots\xRightarrow[]{x} m_1{e}_2^{N}=m_1\xRightarrow[]{x} m_1e_2\xRightarrow[]{x}\cdots,\nonumber\\
   m_2\xRightarrow[]{x} m_2\overline{e}_1\xRightarrow[]{x} m_2\overline{e}_1^2\xRightarrow[]{x} m_2\overline{e}_1^3\xRightarrow[]{x}\cdots\xRightarrow[]{x} m_2\overline{e}_1^{N}=m_2\xRightarrow[]{x} m_2\overline{e}_1\xRightarrow[]{x}\cdots,\label{fs2}
\end{eqnarray}
where equality in the middle comes from the fact that the electric charge is $\mathbb{Z}_N$ and the arrow $\xRightarrow[]{x}$ represents the fractional excitation moving in the $x$-direction by a unit of the lattice spacing.
The trajectory of the magnetic charge is the reminiscent of the form of the dipole of the Wilson loop discussed in the previous subsection~\eqref{dp1} as the intensity of the electric charges attached with the magnetic charge increases one by one when propagating in the horizontal direction~(see also the configuration in the bottom of Fig.~\ref{tj2}). \par
To see this point more explicitly and how the lattice model~\eqref{dp} corresponds to the BF theory with the dipole symmetry~\eqref{foliation dipole}, 
we consider the model~\eqref{dp} on a torus geometry with length being~$L_x$ and~$L_y$ in the~$x$- and $y$-direction to evaluate the GSD. The model~\eqref{dp} admits the following non-contractible Wilson loops in the $x$-direction:~(recall that $\alpha_x=\frac{N}{\gcd(N,L_x)}$)
\begin{eqnarray}
    W_{m_1e_2}(\hy)=\biggl[\prod_{\hx=1}^{L_x}X_{1,\br+\frac{\jy}{2}}Z^{\hx}_{2,\br+\frac{\jx}{2}}\biggr]^{\alpha_x},\quad
    W_{m_2\overline{e}_1}(\hy)=\biggl[\prod_{\hx=1}^{L_x}X_{2,\br+\frac{\jy}{2}}\left(Z_{1,\br+\frac{\jx}{2}}^{\hx}\right)^\dagger\biggr]^{\alpha_x}.
    \label{loopsono1}
\end{eqnarray}
These represent the trajectory of the magnetic charges, dressed with electric charges whose intensity depends on $\hx$. 
Physical interpretation on these loops~\eqref{loopsono1} is that in order for the trajectory of the magnetic charge, which is dressed with electric charges, to form a non-contractible loop, it has to be compatible with the periodic boundary condition in the $x$-direction; the configuration of the charge at $\hx$ and the one at $\hx+L_x$ has to be identical ($\text{mod}\mathbb{Z}_N$). Thus, 
to meet this condition, the magnetic charge has to wind around the torus \textit{multiple times} (more explicitly, $\alpha_x$ times) in the $x$-direction.\par

 Since the fractional statistics of the electric charges is not altered by the arrays of the DW twist terms, we have the following Wilson loops of the electric charges:
\begin{eqnarray}
    W_{e_1}(\hy)=\prod_{\hx=1}^{L_x}X_{1,\br+\frac{\jx}{2}},\quad 
    W_{e_2}(\hy)=\prod_{\hx=1}^{L_x}X_{2,\br+\frac{\jy}{2}}.\label{loopsono2}
\end{eqnarray}
It is straightforward to check these loops~\eqref{loopsono1} and \eqref{loopsono2} are deformable in the $y$-direction, hence $\hy$ independent. 
Since the loops~\eqref{loopsono1} are labeled by the quantum number $(\mathbb{Z} _{\gcd(N,L_x)} )^2$ and the ones~\eqref{loopsono2} by $(\mathbb{Z}_{N} )^2$, the GSD, which amounts to be the number of distinct non-contractible loops of the electric and magnetic charges in the $x$-direction, is given by
\begin{equation}
    {\rm GSD} =[N\times\gcd(N,L_x)]^2 ,
    \label{gsd22}
\end{equation}
which is the identical to~\eqref{gsd} of the BF theory~\eqref{foliation dipole} with dipole symmetry.\par
To summarize the discussion, motivated by the form of the BF theory of a topological phase with dipole symmetry~\eqref{foliation dipole}, we have constructed the corresponding UV lattice model by deciphering the coupling term $A\wedge b\wedge e^x$ as the foliated DW twist terms. In the next subsection, we will see how such DW twist terms affect the statistics of quasi-particles by thoroughly investigating the fusion rules.
\subsection{Alternative derivation of the GSD -- Fusion rules}\label{s33}
One can derive the same GSD of the model~\eqref{gsd22} in a different approach by counting the superselection sectors of the fractional excitations (i.e., the number of distinct fractional excitations) and making use of the property of~\eqref{fs2}, which will be useful for evaluation of GSD in the other topological phase with multipole symmetries.
\par
When we act a single $Z_1$ operator on the ground state at a horizontal link or vertical link, i.e.,~$Z_{1,\br+\frac{\jx}{2}}\ket{\Omega}$, or $Z_{1,\br+\frac{\jy}{2}}\ket{\Omega}$, it induces a pair of $\mathbb{Z}_N$ electric charges, described by the following fusion rule:
\begin{equation}
    I\to e_{1,\br}\otimes\overline{e}_{1,\br+\jx},\quad
    I\to e_{1,\br}\otimes\overline{e}_{1,\br+\jy},
\end{equation}
Here, $I$ denotes the trivial sector. From this fusion rule, one can identify the electric charges which are adjacent with one another, namely,
\begin{equation}
    e_{1,\br}\simeq e_{1,\br+\jx}\simeq e_{1,\br+\jy}\label{fusion1}
\end{equation}
 with $\simeq$ representing the identification of the fractional excitations. Also, due to the periodic boundary condition, we have
\begin{equation}
    e_{1,\br+L_x\jx}\simeq e_{1,\br+L_y\jy}\simeq e_{1,\br}.\label{fusion2}
\end{equation}
The two conditions~\eqref{fusion1} and \eqref{fusion2} imply that
the electric charge $e_{1,\br}$ does not depend on the $x,y$ coordinate, allowing us to set, e.g., $\br=(1,1)\vcentcolon=\br_0$, 
\begin{equation}
    e_{1,\br}\simeq e_{1,\br_0}(\vcentcolon=\mathfrak{E}_{1}).
\end{equation}
Likewise, it can be shown that the electric charge $e_{2,\br}$ does not depend on the $x,y$ coordinate, and one can set $\br=\br_0$, 
\begin{equation}
      e_{2,\br}\simeq e_{2,\br_0}(\vcentcolon=\mathfrak{E}_{2}).\label{pp}
\end{equation}
Therefore, the theory~\eqref{dp} admits $(\mathbb{Z}_N )^2$ electric charges, labeled by $\mathfrak{E}_{1}^{u}\mathfrak{E}_{2}^{v}\;(u,v\in\mathbb{Z}_N)$.
\par
When it comes to the magnetic charges, we think of applying a single $X_1$ operator on the ground state at a horizontal or vertical link, that is, $X_{1,\br+\frac{\jx}{2}}\ket{\Omega}$, or $X_{1,\br+\frac{\jy}{2}}\ket{\Omega}$. In the former case, we have the following fusion rule
\begin{equation}
    I\to  m_{1,\bp}\otimes\overline{m}_{1,\bp-\jy},\label{fs33}
\end{equation}
whereas in the latter case, we find
\begin{equation}
    I\to{m}_{1,\bp-\jx}\otimes  \overline{m}_{1,\bp}\otimes \overline{e}_{2,\br}.\label{fs44}
\end{equation}
The fusion rule~\eqref{fs44} is the concise description of what we have seen in the previous section, that is, when a magnetic charge goes in the $x$-direction by a unit of the lattice spacing, an electric charge is attached to it~\eqref{fs2}.
From the fusion rule~\eqref{fs33}, 
we have
\begin{equation}
      m_{1,\bp}\simeq{m}_{1,\bp-\jy}.
\end{equation}
From this relation and the periodic boundary condition in the $y$-direction, $m_{1,\bp+L_y\jy}\simeq m_{1,\bp}$, it follows that the magnetic charge $m_{1,\bp}$ does not depend on the $y$-coordinate, allowing us to set~$\hy=1$, that is, $\bp=(\hx+\fr,\frac{3}{2})\vcentcolon=\mathbf{q}_0$.
Also, from the fusion rule~\eqref{fs44}, jointly with~\eqref{pp}, one finds
\begin{equation}
  {m}_{1,\mathbf{q}_0}\simeq m_{1,\mathbf{q}_0-\jx}\otimes {\overline{\mathfrak{E}}}_{2}.\label{fs3}
\end{equation}
The adjacent magnetic charges in the $x$-direction are identified up to an electric charge. Similarly to the argument in the $y$-direction, one naively wonders the magnetic charge also does not depend on $x$ coordinate. However, we have to carefully take into account the periodic boundary condition in the $x$-direction and the effect of the electric charge which is attached to the magnetic charge. Indeed,
by iterative use of~\eqref{fs3}, we have 
\begin{equation}
     m_{1,\mathbf{q}_0+K_x\jx}\simeq m_{1,\mathbf{q}_0}\otimes(\overline{{\mathfrak{E}}_{2}})^{K_x}\label{fs4}\;(K_x\in\mathbb{Z}).
\end{equation}
The condition~\eqref{fs4} needs to be compatible with the periodic boundary condition in the $x$-direction.
    Hence, depending on $N$ and $L_x$, the admissible magnetic excitation is the \textit{multiple} magnetic charges rather than a single magnetic charge. We think of the multipole magnetic charges, denoted by $m^{\alpha_x}_{1,\mathbf{q}_0}$, with $\alpha_x=\frac{N}{\gcd(N,L_x)}$, which satisfies
    \begin{equation}
        {m}^{\alpha_x}_{1,\mathbf{q}_0}\simeq m^{\alpha_x}_{1,\mathbf{q}_0-\jx}\otimes \overline{{\mathfrak{E}}}^{\alpha_x}_{2}\label{ki}
    \end{equation}
    and the periodic boundary condition
    \begin{equation}
          m^{\alpha_x}_{1,\mathbf{q}_0+L_x\jx}\simeq  m^{\alpha_x}_{1,\mathbf{q}_0}.\label{ki2}
    \end{equation}
    Hence, from these two conditions~\eqref{ki} and \eqref{ki2}, it follows that the magnetic charge $m^{\alpha_x}_{1,\mathbf{q}_0}$ at any $x$ coordinate is generated by 
an electric charge $\mathfrak{E}_2$ and [$\bp_0\vcentcolon=(\frac{3}{2},\frac{3}{2})$]
    \begin{equation}
        m^{\alpha_x}_{1,\bp_0}(\vcentcolon=\mathfrak{M}_1),
    \end{equation}
  indicating that the magnetic charge induced by the Pauli operators $X_1'$s, is labeled by the $\mathfrak{M}_1$.
Analogous analysis leads us to that the magnetic charge induced by the Pauli operators $X_2'$s, is labeled by
\begin{equation}
    m^{\alpha_x}_{2,\bp_0}(\vcentcolon=\mathfrak{M}_2).
\end{equation}
Therefore, the theory supports $(\mathbb{Z}_{\gcd(N,L_x)} )^2$ magnetic charges, which are characterized by $(\mathfrak{M}_1)^w(\mathfrak{M}_2)^s$ with $w,s\in\mathbb{Z}_{\gcd(N,L_x)}$.\par
In summary, via a close investigation of the fusion rules, we have evaluated the number of superselection sectors of the model~\eqref{dp}. There are $N^2$ electric charges and $\gcd(N,L_x)^2$ magnetic charges. Hence, the GSD on the torus, which amounts to be the number of the superselection sectors, is given by
${\rm GSD}=[N\times\gcd(N,L_x)]^2$, which is the same as~\eqref{gsd22}.

\section{Dipole symmetries in $x$- and $y$-direction}\label{s4}
\begin{figure}[t]
    \begin{center}
      \begin{subfigure}[h]{0.14\textwidth}
  \includegraphics[width=\textwidth]{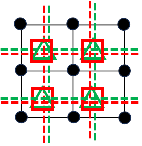}
         \caption{}\label{three}
             \end{subfigure}
                 \begin{subfigure}[h]{0.20\textwidth}
  \includegraphics[width=\textwidth]{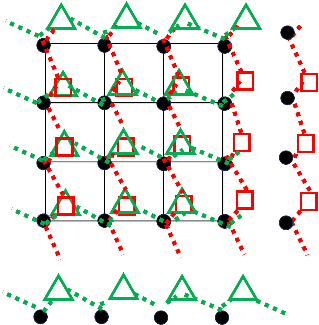}
         \caption{}\label{czgate}
             \end{subfigure}
                              \begin{subfigure}[h]{0.34\textwidth}
  \includegraphics[width=\textwidth]{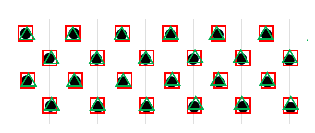}
         \caption{}\label{modelxy}
             \end{subfigure}
                  \begin{subfigure}[h]{0.84\textwidth}
  \includegraphics[width=\textwidth]{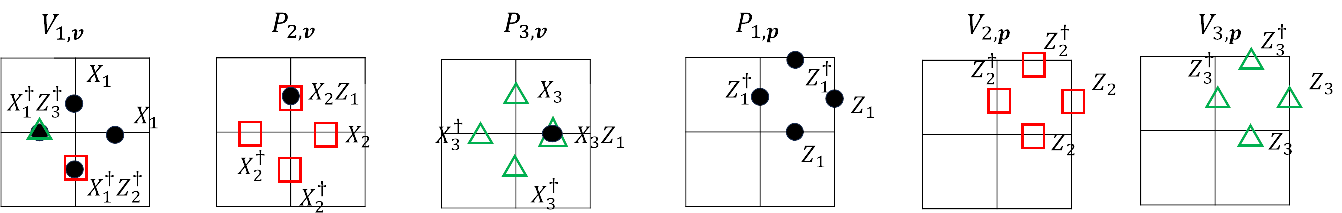}
         \caption{}\label{VP}
             \end{subfigure}
                               \begin{subfigure}[h]{0.34\textwidth}
  \includegraphics[width=\textwidth]{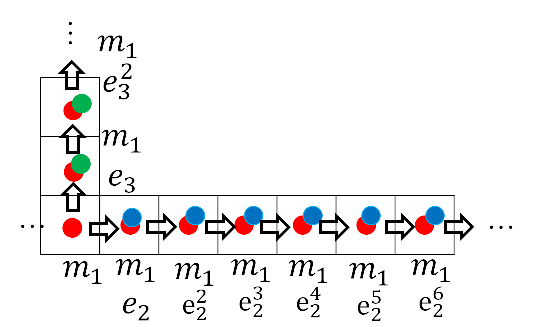}
         \caption{}\label{fs444}
             \end{subfigure}
               \end{center}
\caption{(a) Three 2D square lattices. Two of them are introduced so that their nodes (red squares and green triangles) are located at the center of the plaquette of the other lattice (black nodes and links). 
(b)~CZ gates~\eqref{cz3} are implemented along the red and green zigzag path. 
(c) After gauging the global $\mathbb{Z}_N$ symmetry, gauge fields are located on each link (black dot, red square, and green triangle). (d) The terms defined on the lattice~(c) that constitutes the Hamiltonian~\eqref{death1} and \eqref{death2}. 
(e) Schematic illustration how the magnetic charge $m_1$ is dressed with electric charges when it goes in the $x$- or $y$-direction. }
   \end{figure}
One can generalize the previous argument given in sec.~\ref{2.2} to the case where we prepare stacking of the SPT phases in the $y$-direction as well as the $x$-direction. 
Consider three layers of 2D square lattices,
The orientation of the links is the same as the previous case~(Fig.~\ref{sfa}). 
We portray the lattices in Fig.~\ref{three} with intentionally shifting two of them with respect to the other for visual illustration.
We define the Hilbert space of the $\mathbb{Z}_N$ qubit on each node of the lattices as $\ket{a}_i$ $(a\in\mathbb{Z}_N )$ with index $i$ distinguishing the three lattices. 
Pictorially $i=1,2$ and 3 in Fig.~\ref{three} correspond to the lattice with black, red and green colors respectively. 
Then we introduce the Hamiltonian of the paramagnet as
\begin{equation}
  \widetilde{H}_0=-\sum_{\br}\left(\tau^X_{1,\br}+\tau^X_{2,\br}+\tau^X_{3,\br}\right).\label{hami02}
\end{equation}
Here, $\tau_{i,\br}^{X}$ is the spin flip operator at a node of the three lattices.
We implement the CZ gates on the lattices along the $x$ (green)- and $y$ (red)-direction in accordance with the zigzag path in Fig.~\ref{czgate}. More explicitly, we act the following operator consisting of the CZ gates
\begin{eqnarray}
     \widetilde{U}_{CZ}\vcentcolon=\prod_{\br}CZ^\dagger_{(1,\br),(2,\br)}CZ_{(1,\br),(2,-\jy)}
     \times CZ^\dagger_{(1,\br),(3,\br)}CZ_{(1,\br),(3,\br-\jx)}.\label{cz3}
\end{eqnarray}
on the Hamiltonian~\eqref{hami02} which yields
\begin{eqnarray}
      \widetilde{U}^{\dagger}_{CZ}  \widetilde{H} _0 \widetilde{U}_{CZ}
   =-\sum_{\br}\biggl[(\tau^Z)^\dagger_{3,\br-\jx}(\tau^Z)^\dagger_{2,\br-\jy}\tau^X_{1,\br}\tau^Z_{2,\br}\tau^Z_{3,\br}+(\tau^Z)^\dagger_{1,\br+\mathbf{j}_y}\tau^X_{2,\br}\tau^Z_{1,\br}+Z^\dagger_{1,\br+\mathbf{j}_x}\tau^X_{3,\br}\tau^Z_{1,\br}\biggr]. \nonumber\\
      \label{shine}
\end{eqnarray}
The Hamiltonian~\eqref{shine} respects $(\mathbb{Z}_N )^3$ global symmetry, corresponding to global spin flip on the three layers of the lattices,
 \begin{equation*}
  \prod_{\br}\tau^X_{i,\br} \quad  (i=1,2,3) ,
\end{equation*}
which we can gauge. Following the same procedure explained in the previous section (sec.~\ref{2.2}),
gauged Hamiltonian is given by
\begin{equation}
    \widetilde{H}_{\rm dipole}
    =-\biggl[\sum_{\br}{V}_{1,\br}+\sum_{\bp}{P}_{1,\bp}\biggr] -\biggl[\sum_{\br}{V}_{2,\br}+\sum_{\bp}{P}_{2,\bp}\biggr] -\biggl[\sum_{\br}{V}_{3,\br}+\sum_{\bp}{P}_{3,\bp}\biggr] +h.c.
    \label{gauged4}
\end{equation}
with
\begin{eqnarray}
    {V}_{1,\br} &\vcentcolon=& Z^\dagger_{3,\br-\frac{\jx}{2}}Z^\dagger_{2,\br-\frac{\jy}{2}}X_{1,\br+\frac{\jx}{2}}X_{1,\br+\frac{\jy}{2}}X^{\dagger}_{1,\br-\frac{\jx}{2}}X^{\dagger}_{1,\br-\frac{\jy}{2}},\nonumber\\
  {V}_{2,\br} &\vcentcolon=& Z_{1,\br+\frac{\jy}{2}}X_{2,\br-\frac{\jy}{2}}^\dagger X_{2,\br-\frac{\jx}{2}}^\dagger X_{2,\br+\frac{\jx}{2}}X_{2,\br+\frac{\jy}{2}},\nonumber\\
  {V}_{3,\br} &\vcentcolon=& Z_{1,\br+\frac{\jx}{2}}X_{3,\br-\frac{\jy}{2}}^\dagger X_{3,\br-\frac{\jx}{2}}^\dagger X_{3,\br+\frac{\jx}{2}}X_{3,\br+\frac{\jy}{2}},\label{death1}
\end{eqnarray}
and
\begin{eqnarray}
    P_{i,\bp}\vcentcolon=Z_{i,\bp-\frac{\jy}{2}}Z_{i,\bp+\frac{\jx}{2}}Z^{\dagger}_{i,\bp+\frac{\jy}{2}}Z^\dagger_{i,\bp-\frac{\jx}{2}} ,
  \label{death2}
\end{eqnarray}
where $X_{i}/Z_{i}$ represents the Pauli operator of the gauge field which reside on the links (Fig.~\ref{modelxy}).
These terms are depicted in Fig.~\ref{VP}. The Hamiltonian~\eqref{gauged4} with~\eqref{death1} and \eqref{death2} resembles three copies of the~$\mathbb{Z}_N$ toric codes with the difference being that a few Pauli operators are multiplied with the vertex terms~\eqref{death1}, which comes from the foliated DW twist terms. The ground state of the model~\eqref{gauged4} is the projected state~$\ket{\Omega}$, satisfying
\begin{eqnarray}
      {V}_{i,\br}\ket{\Omega}=\ket{\Omega},\quad 
      P_{i,\bp}\ket{\Omega}=\ket{\Omega} \qquad \forall i,\br,\bp .
      \label{condi}
\end{eqnarray}
\par
By the similar argument in Sec.~\ref{s33}, one can discuss how behavior of the fractional excitations is affected by the foliated DW twist terms. To this end, we denote a fractional electric charge which violates the first condition in~\eqref{condi} with eigenvalue $\omega$ by $e_{i,\br}$. Similarly, we label a fractional magnetic charge which violates the second condition given in~\eqref{condi} with eigenvalue $\omega$ by 
$m_{i,\bp}$.\par
It can be shown that no additional charge is attached to electric charges when they move. Regarding the magnetic charges, depending on the direction they go, electric charges are attached to them. For instance, by the same logic given in around~\eqref{fs2}, one can see that the magnetic charge $m_{1}$ (we omit the coordinate for simplicity) is dressed with an electric charge $e_2$ ($e_3$) every time it travels in the $x(y)$-direction, schematically described by (see also Fig.~\ref{fs444})
\begin{eqnarray}
      m_1 \xRightarrow[]{x} m_1e_2\xRightarrow[]{x} m_1e_2^2\xRightarrow[]{x} m_1e_2^3\xRightarrow[]{x}\cdots\xRightarrow[]{x} m_1{e}_2^{N}=m_1\xRightarrow[]{x} m_1e_2\xRightarrow[]{x}\cdots,\nonumber\\
   m_1\xRightarrow[]{y} m_1{e}_3\xRightarrow[]{y} m_1{e}_3^2\xRightarrow[]{y} m_1{e}_3^3\xRightarrow[]{y}\cdots\xRightarrow[]{y} m_1{e}_3^{N}=m_1\xRightarrow[]{y} m_1{e}_3\xRightarrow[]{y}\cdots.\label{kk}
\end{eqnarray}
Here $\xRightarrow[]{x/y}$ indicates the fractional excitation goes in the $x/y$-direction by a unit of the lattice spacing. By the same token, one finds that whenever the magnetic charge $m_2$ ($m_3$) goes in the $x$($y$)-direction, an electric charge is attached to it, described by 
\begin{eqnarray}
      m_2 \xRightarrow[]{x} m_2\overline{e}_1\xRightarrow[]{x} m_2\overline{e}_1^2\xRightarrow[]{x} m_2\overline{e}_1^3\xRightarrow[]{x}\cdots\xRightarrow[]{x} m_2\overline{e}_1^{N}=m_2\xRightarrow[]{x} m_2\overline{e}_1\xRightarrow[]{x}\cdots,\nonumber\\
   m_3\xRightarrow[]{y} m_3\overline{e}_1\xRightarrow[]{y} m_3{e}_1^2\xRightarrow[]{y} m_3\overline{e}_1^3\xRightarrow[]{y}\cdots\xRightarrow[]{y} m_3\overline{e}_1^{N}=m_3\xRightarrow[]{y} m_3\overline{e}_1\xRightarrow[]{y}\cdots.
\end{eqnarray}
Also, one can verify that no additional charge is attached to the magnetic charge $m_2$ ($m_3$) when it travels in the $y$-($x$-)direction.

We can evaluate the GSD of this model on the torus geometry by investigating the superselection sectors of the fractional excitations, analogously to~Sec.~\ref{s33}. 
By considering the fusion rules of the electric charges, obtained by acting the Pauli operator $Z_1$, $Z_2$ and $Z_3$ on the ground state, jointly with the periodic boundary condition in the $x$- and $y$-direction, one finds that the electric charges, $e_{i,\br}$ at any coordinate of $\hx$ and $\hy$, are labeled by
\begin{equation}
    e_{1,\br}\simeq  e_{i,\br_0}(\vcentcolon=\mathfrak{E}_i) , 
    \label{hey3}
\end{equation}
where $i=1,2,3$ and $\br_0=(1,1)$.
Thus, there are $N^3$ electric charges. As for the magnetic charges, by acting the Pauli operator $X_1$ on the ground state at the vertical link or the horizontal link, we have the following fusion rules.
\begin{eqnarray}
    I\to m_{1,\bp-\jx}\otimes \overline{m}_{1,\bp}\otimes \overline{e}_{2,\br},\quad
      I\to \overline{m}_{1,\bp-\jy}\otimes {m}_{1,\bp}\otimes e_{3,\br},
\end{eqnarray}
which is the concise statement of~\eqref{kk}. From these fusion rules, jointly with~\eqref{hey3}, we have
\begin{equation}
      {m}_{1,\bp}\simeq m_{1,\bp-\jx}\otimes \overline{{\mathfrak{E}}}_{2},\quad  {m}_{1,\bp}\simeq m_{1,\bp-\jy}\otimes \overline{\mathfrak{E}}_{3}.
      \label{hey4}
\end{equation}
The iterative use of~\eqref{hey4} gives
\begin{equation}
      {m}_{1,\bp+K_x\jx+K_y\jy}
      \simeq   {m}_{1,\bp}\otimes \overline{{\mathfrak{E}}}_{2}^{K_x}\otimes\overline{\mathfrak{E}}_{3}^{K_y}\quad
      (K_x,K_y\in\mathbb{Z}).
      \label{mantle}
\end{equation}
In order for~\eqref{mantle} to be consistent with the periodic boundary condition, the multiple of magnetic charges ${m}^{\alpha_{xy}}_{1,\bp}$ rather than a single one are allowed. 
Here, $\alpha_{xy}\vcentcolon=\frac{N}{\gcd(N,L_x,L_y)}$. 
One can verify that the charge~${m}^{\alpha_{xy}}_{1,\bp}$ at any coordinate of $\hx$ and $\hy$ is generated by $\overline{{\mathfrak{E}}}_{2}$, $\overline{\mathfrak{E}}_{3}$, and
\begin{equation}
    {m}^{\alpha_{xy}}_{1,\bp_0}(\vcentcolon={\mathfrak{M}}_{1})\quad
    {\rm with}\ \ \bp_0=\left(\frac{3}{2},\frac{3}{2}\right) ,
\end{equation}
which indicates that the magnetic charge induced by applying the Pauli operator $X_1$ on the ground state is characterized by $(\mathfrak{M}_{1})^{s_1}$ with $s_1\in\mathbb{Z}_{\gcd(N,L_x,L_y)}$. Similar discussion leads to that the magnetic charge induced by applying the Pauli operator $X_2$ and $X_3$ on the ground state is labeled by
\begin{equation}
     {m}^{\alpha_{x}}_{2,\bp_0}(\vcentcolon={\mathfrak{M}}_{2}) \quad {\rm and}\quad
     {m}^{\alpha_{y}}_{3,\bp_0}(\vcentcolon={\mathfrak{M}}_{3}),
\end{equation}
respectively, with $\alpha_x=\frac{N}{\gcd(N,L_x)}$ and $\alpha_y=\frac{N}{\gcd(N,L_y)}$.
Overall, there are in total $\gcd(N,L_x,L_y)\times\gcd(N,L_x)\times\gcd(N,L_y)$ magnetic charges.
\par
To recap the argument, by investigating the fusion rules of the fractional excitations, we identify the number of the superselection sectors of the model on the torus geometry. The excitation is labeled by
\begin{equation}
( \mathfrak{E}_{1})^{w_1}  ( \mathfrak{E}_{2})^{w_2}  (\mathfrak{E}_{3})^{w_3}    (\mathfrak{M}_{1})^{s_1}   (\mathfrak{M}_{2})^{s_2}   (\mathfrak{M}_{3})^{s_3}\;\;
\end{equation}
with $w_i\in \mathbb{Z}_N\;(i=1,2,3)$, $s_1\in\mathbb{Z}_{\gcd(N,L_x,L_y)}$, $s_2\in\mathbb{Z}_{\gcd(N,L_x)}$ and $s_3\in\mathbb{Z}_{\gcd(N,L_y)}$. 
Thus, the number of the superselection sectors, which is equivalent to the GSD on the torus, is given by
\begin{equation}
    {\rm GSD} =N^3\times\gcd(N,L_x,L_y)\times\gcd(N,L_x)\times\gcd(N,L_y).\label{gsd3}
\end{equation}
The model that we have considered in this section corresponds to the following BF theory~\cite{2023arXiv231006701E}:
\begin{equation}
  \mathcal{L}_{ {\rm dipole}-x,y}=\frac{N}{2\pi}a\wedge db+\sum_{I=x,y}\frac{N}{2\pi}A^I\wedge dc^I+\frac{N}{2\pi}A^I\wedge b\wedge e^I.\label{foliation dipole_xy}
\end{equation}
Here, the field $a,b,c^I,A^I$ denotes a $U(1)$ $1$-form gauge field. Also, $e^I$ describes the foliation field, i.e, $e^x=dx,e^y=dy$.
The gauge field $a$, $A^x$, and $A^y$ is associated with global and dipole symmetry in the $x$- and $y$-direction, respectively. The BF theory~\eqref{foliation dipole_xy} shows the same GSD as~\eqref{gsd3}~\cite{2023arXiv231006701E}.
Also, it is interesting to note that a stabilizer model in a different form obtained by Higgssing the tensor gauge theory~\cite{tensor_gauge}, which is a generalized Maxwell theory preserving dipole charges, exhibits the same GSD as~\eqref{gsd3}~\cite{PhysRevB.106.045145,oh2022rank}.
 
\section{Construction of other fracton models}
\label{s5}
  So far we have demonstrated a concrete lattice realization of the topological phases with multipole symmetries. 
  One can apply the approach discussed in the previous sections to other fracton model, such as the $\mathbb{Z}_N$ exotic theory and
 $X$-cube model. 

\subsection{Exotic $\mathbb{Z}_N$ theory}
To start, let us first introduce BF theory description of the fracton model, which can be obtained by the BF theory of topological phases with multipole symmetries with a slight modification.
If we replace the gauge field~$c^I$ with $\phi^I e^I$ ($I$ is not summed over) in~\eqref{foliation dipole_xy}, where $\phi^I$ is a $0$-form field, we have
\begin{equation}
 \mathcal{L}_{EX}=\frac{N}{2\pi}a\wedge db+\sum_{I=x,y}\frac{N}{2\pi}A^I\wedge d\phi^I\wedge e^I+\frac{N}{2\pi}A^I\wedge b\wedge e^I,\label{foliation ex}
   \end{equation}
which describes the exotic $\mathbb{Z}_N$ gauge theory~\cite{seiberg2021exotic}.
Compared with~\eqref{foliation dipole_xy}, which has the dipole symmetries, theory~\eqref{foliation ex} possesses an additional symmetry: the theory is invariant under $A^I\to A^I+e^Ig$ with $g$ being an arbitrary function. 
This symmetry is known as \textit{subsystem symmetry}~\cite{PhysRevB.66.054526,Seiberg:2019vrp,Vijay}, which is of importance in the context of the fracton topological phases. A theory with such a symmetry admits a symmetry operation on a sub-manifold rather than entire space. 
Manifestation of the subsystem symmetry in the theory~\eqref{foliation ex} is that a gauge invariant operator constructed by the gauge field $A^I$ is mobile only along a sub-manifold, 
which forbids moving in the direction parallel to~$e^I$.
\par
The last term in~\eqref{foliation ex} has the exactly the same form as the one we have encountered in the previous sections, the foliated DW twist term. Also, the second term that enters~\eqref{foliation ex} describes foliated topological field theories each of which corresponds to a spontaneously symmetry breaking phase~\cite{kapustin2014coupling}. Based on this observation, it would be tantalizing to implement the similar procedure as we did in the previous argument to construct the exotic $\mathbb{Z}_N$ gauge theory via gauging SPT phases.

 \begin{figure}[t]
    \begin{center}
      \begin{subfigure}[h]{0.15\textwidth}
  \includegraphics[width=\textwidth]{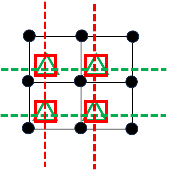}
         \caption{}
         \label{ex1}
             \end{subfigure}
                 \begin{subfigure}[h]{0.55\textwidth}
  \includegraphics[width=\textwidth]{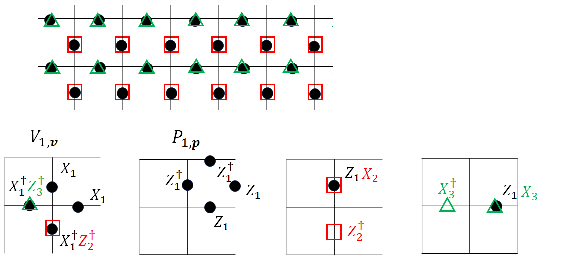}
         \caption{}
         \label{ex2}
             \end{subfigure}
                \begin{subfigure}[h]{0.49\textwidth}
  \includegraphics[width=\textwidth]{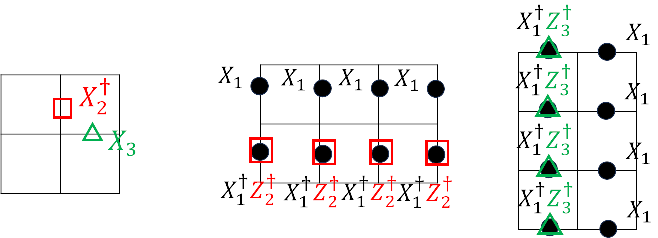}
         \caption{}
         \label{ex3}
             \end{subfigure}
               \end{center}
\caption{(a)~2D square lattice and arrays of 1D chains stacked along the $x$-direction (red squares and dashed lines) and the ones along the $y$-direction (green triangles and dashed lines). We set the orientation of the links in the same way as the previous cases. (b) Top: The lattice configuration after gauging the global~$\mathbb{Z}_N$ symmetry where gauge fields reside on links. Bottom: Terms that constitute the Hamiltonian~\eqref{d33}. (c) Examples of the logical operators given in~\eqref{ope}. }
\end{figure}

%
To this end, we think of a 2D square lattice jointly with arrays of 1D chains stacked along both of $x$- and $y$-directions, as demonstrated in Fig.~\ref{ex1} (note the distinction between the configurations in Fig.~\ref{ex1} and Fig.~\ref{three}.) with periodic boundary condition being imposed~(i.e., $\hx\sim\hx+L_x$, $\hy\sim\hy+L_y$).
We define the Hilbert space of $\mathbb{Z}_N$ qubit on a node of the lattice by $\ket{a}_i$ with $i$ distinguishing the 2D lattice ($i=1$ black dot in Fig.~\ref{ex1}) and that of the 1D chain stacked along $x$- ($i=2$ red square in Fig.~\ref{ex1}) and $y$-direction ($i=3$ green triangle in Fig.~\ref{ex1}).
We also introduce Hamiltonian of this lattice as
\begin{equation}
     \widetilde{H}_0=-\sum_{\br}\left(\tau^X_{1,\br}+\tau^X_{2,\br}+\tau^X_{3,\br}\right).\label{hami3}
\end{equation}
Even though the Hamiltonian has the same form as~\eqref{hami02}, there is a crucial difference between the two; in the present case there is no link connecting between the states $\ket{a}_{2,\br}$ and $\ket{a}_{2,\br\pm\jx}$ in the horizontal direction. Likewise, there is no link that connects the states $\ket{a}_{3,\br}$ and $\ket{a}_{3,\br\pm\jy}$ in the vertical direction.\par
We implement the same CZ gates~\eqref{cz3} on the Hamiltonian~\eqref{hami3}, yielding
 \begin{eqnarray}
   \widetilde{U}^{\dagger}_{CZ}  \widetilde{H} _0 \widetilde{U}_{CZ} =-\sum_{\br}\biggl[(\tau^Z)^\dagger_{3,\br-\jx}(\tau^Z)^\dagger_{2,\br-\jy}\tau^X_{1,\br}\tau^Z_{2,\br}\tau^Z_{3,\br}+(\tau^Z)^\dagger_{1,\br+\mathbf{j}_y}\tau^X_{2,\br}\tau^Z_{1,\br}+Z^\dagger_{1,\br+\mathbf{j}_x}\tau^X_{3,\br}\tau^Z_{1,\br}\biggr].
   \nonumber\\  \label{shine1}
\end{eqnarray}
The model has the one $\mathbb{Z}_N$ global and two $\mathbb{Z}_N$ subsystem symmetries generated by 
 \begin{eqnarray}
     \prod_{\br}\tau^X_{1,\br},\quad 
     \prod_{\hy=1}^{L_y}\tau^X_{2,\br}\;\;(1\leq \hx\leq L_x),\quad
     \prod_{\hx=1}^{L_x}\tau^X_{3,\br}\;\;(1\leq \hy\leq L_y).
 \end{eqnarray}
 Following the same procedure outlined around~\eqref{spt3}, one can gauge these symmetries. 
 The gauged Hamiltonian reads (see also Fig.~\ref{ex2})
 \begin{eqnarray}
     H_{EX}=&-&\biggl[\sum_{\br}{V}_{1,\br}+\sum_{\bp}{P}_{1,\bp}\biggr]\nonumber\\
     &-&\sum_{\hx}H_{fr,x}(\hx)-\sum_{\hy}H_{Fr,y}(\hy)+h.c.~.
     \label{d33}
 \end{eqnarray}
 The first line describes the $\mathbb{Z}_N$ toric code consisting of the terms given in~\eqref{death1} and \eqref{death2} whereas the second line corresponds to ferromagnetic chains stacked along $x$- and $y$-direction:
 \begin{eqnarray}
     H(\hx)_{Fr,x}=\sum_{\hy}Z_{1,\br+\jy}X_{2,\br}X^\dagger_{2,\br-\jy},\quad
     H(\hy)_{Fr,y}=\sum_{\hx}Z_{1,\br+\jx}X_{3,\br}X^\dagger_{3,\br-\jx} .
 \end{eqnarray}
  Note that due to the DW twist terms obtained by gauging the foliated SPT phases, an additional Pauli operator $Z_i$ is attached to the terms (see also Fig.~\ref{ex3}).
The ground state of the model~\eqref{d33} is the projected state, satisfying 
\begin{eqnarray}
&&    {V}_{1,\br}\ket{\Omega}= {P}_{1,\bp}\ket{\Omega}=\ket{\Omega}\quad \forall\br,\bp ,\nonumber\\
&& Z_{1,\br+\jy}X_{2,\br}X^\dagger_{2,\br-\jy}\ket{\Omega}=\ket{\Omega},\quad 
 Z_{1,\br+\jx}X_{3,\br}X^\dagger_{3,\br-\jx}\ket{\Omega}=\ket{\Omega}\quad \forall\br .
 \label{omega}
\end{eqnarray}
\par
One can evaluate the GSD of the model on the torus geometry. The model~\eqref{d33} admits the following logical operators i.e., operators that commute with the Hamiltonian~\eqref{d33}:
\begin{eqnarray}
    \zeta_{\br} &\vcentcolon=& X_{3,\br+\frac{\jx}{2}}X^\dagger_{2,\br+\frac{\jy}{2}},\nonumber\\
    \eta_{x}(\hy)&\vcentcolon=& \prod_{\hx=1}^{L_x}Z_{2,\br-\frac{\jy}{2}}X^\dagger_{1,\br-\frac{\jy}{2}}X_{1,\br+\frac{\jy}{2}}\quad (1\leq \hy\leq L_y), \nonumber\\
    \;\eta_{y}(\hx)&\vcentcolon=& \prod_{\hy=1}^{L_y}Z_{3,\br-\frac{\jx}{2}}X^\dagger_{1,\br-\frac{\jx}{2}}X_{1,\br+\frac{\jx}{2}}\quad (1\leq \hx\leq L_x).\label{ope}
\end{eqnarray}
Examples of these operators are shown in Fig.~\ref{ex3}. 
The form of the last two operators resemble the Wilson loop introduced in the exotic $\mathbb{Z}_N$ theory constructed by a non-contractible loop of a symmetric tensor gauge field~\cite{seiberg2021exotic}.
There are a few constrains on the operators~\eqref{ope}. To see this, we multiply 
the term $V_{1,\br}^\dagger$ on entire vertex gives
\begin{eqnarray}
   1= \prod_{\br}V_{1,\br}^\dagger=\prod_{\hy}\eta_{x}(\hy)\times\prod_{\hx}\eta_{y}(\hx),\label{con}
\end{eqnarray}
where we have used~\eqref{omega}. Also, by combination of the terms that enter the Hamiltonian~\eqref{d33}, together with~\eqref{omega} gives
\begin{eqnarray}
    1= \zeta_{\br} \zeta^\dagger_{\br+\frac{\jx}{2}} \zeta^\dagger_{\br+\frac{\jy}{2}} \zeta_{\br+\frac{\jx}{2}+\frac{\jy}{2}}.\label{con2}
\end{eqnarray}
These constraints~\eqref{con} and \eqref{con2} indicate that there are $L_x+L_y-1$ distinct number of the logical operators~$\zeta_{\br} $, and the same number of distinct logical operators $\eta_{x}(\hy)$ and $\eta_{y}(\hx)$. One can show that pairs of such logical operators, each of which generates $\mathbb{Z}_N$ Heisenberg algebra gives the GSD as
 \begin{equation}
    {\rm GSD} =N^{L_x+L_y-1}.
 \end{equation}
This is in agreement with~\cite{seiberg2021exotic}.

\subsection{X-cube model}
One can analogously establish the X-cube model in our approach.
To this end, we recall the BF theory description of the X-cube model~\cite{foliated1}. 
Consider the following BF theory defined in $(3+1)$d
  \begin{equation}
        \mathcal{L}_{ {\rm dipole}-xyz}
        =\frac{N}{2\pi}a\wedge db+\sum_{I=x,y,z}\frac{N}{2\pi}A^I\wedge dc^I+\frac{N}{2\pi}A^I\wedge b\wedge e^I,\label{foliation dipole_xyz}
  \end{equation}
where the fields $a,A^I$ ($b,c^I$) represent $U(1)$ $1$-form ($2$-form) gauge fields and $e^x=dx, e^y=dy,e^z=dz$. This BF theory is the $(3+1)$d analogue of~\eqref{foliation dipole_xy}, where the gauge field $a,A^I$ is associated with global and dipole in the $I$th direction, respectively. 
We replace the $2$-form gauge field $c^I$ with $dB^I\wedge e^I$ ($I$ is not summed over) where $B^I$ denotes a $U(1)$ $1$-form gauge field. With this replacement, the BF theory~\eqref{foliation dipole_xyz} becomes
   \begin{equation}
 \mathcal{L}=\frac{N}{2\pi}a\wedge db+\sum_{I=x,y,z}\frac{N}{2\pi}A^I\wedge dB^I\wedge e^I+\frac{N}{2\pi}A^I\wedge b\wedge e^I,\label{xc}
\end{equation}
which is known as the effective field theory of the X-cube model~\cite{foliated1}. Similarly to~\eqref{foliation ex}, the theory has the subsystem symmetry; the theory is invariant under $A^I\to A^I+ge^I$.

Note that the second term in~\eqref{xc} describes the 2D toric codes, forming stacked layers and also that 
the last term in~\eqref{xc} has the same form as the one we have discussed so far, allowing us to carry out our approach to build up the UV model which corresponds to the field theory~\eqref{xc}. We emphasize that although realization of the X-cube model via gauging the sub-system symmetry was discussed previously~\cite{Vijay}, 
we believe that it is still useful to present our way to construct the X-cube model via gauging the foliated SPT phases. 
This is because the way we construct the model has a broad perspective in the sense that it is applicable to both of topological phases with multipole symmetries and other fracton models, and allows us to make a transparent understanding of the topological field theory of these phases by interpreting the coupling term $A^I\wedge b\wedge e^I$ as the foliated DW twist term. 
Furthermore, we believe that our approach may find an application to realization of the fracton topological phases with more generic group symmetries including non-Abelian, as our approach complies well with the group cohomology that describes SPT phases. 
We leave more details of these issues for future studies.
\par

  \begin{figure}[t]
    \begin{center}
      \begin{subfigure}[h]{0.35\textwidth}
  \includegraphics[width=\textwidth]{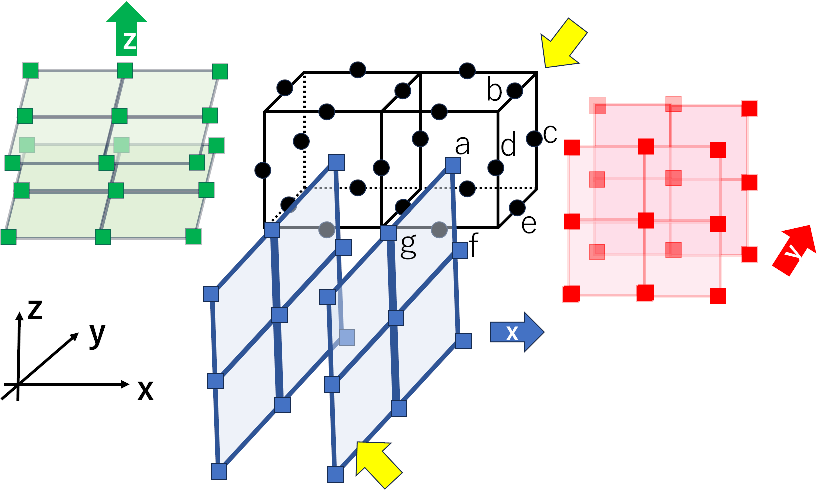}
         \caption{}\label{xcube1}
             \end{subfigure}
                 \begin{subfigure}[h]{0.40\textwidth}
  \includegraphics[width=\textwidth]{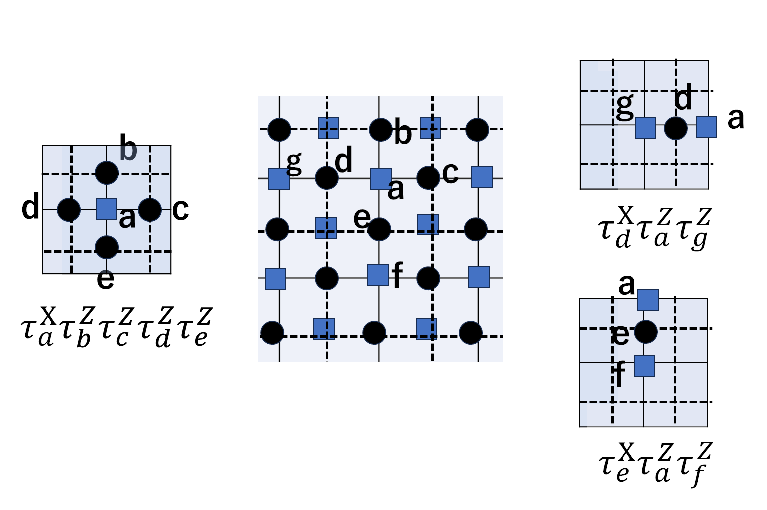}
         \caption{}\label{xcube2}
             \end{subfigure}
                         \end{center}
\caption{(a) 3D cubic lattice and layers of 2D lattices. Each node of the 2D lattice is located at the center of the cube. (b) Side view of the lattice from the $x$-axis is shown in the middle large panel. The face of the cubic lattice and the layer of 2D lattice indicated by the yellow arrow in~(a) constitute a 2D lattice. The CZ gates are implemented between states on the the layer of the 2D lattices and the ones on the face of the cube. Example of the CZ gates are depicted in the small panels on the left and right.
The node with alphabet corresponds to the one in~(a). 
}
   \end{figure}

With these regards in mind, now we turn to constructing the X-cube model via gauging foliated SPT phases in the case of $N=2$ with system size $L_x\times L_y\times L_z$ for simplicity.
To do this, we envisage a cubic lattice and layers of the 2D lattices stacked in the $x$, $y$, and $z$-directions, which corresponds to blue, red, and green layers in Fig.~\ref{xcube1} respectively. 
Each stacked layers are separated by a unit lattice spacing and each node of the 2D lattice is located at the center of the cube. 
We introduce qubits on each link of the cubic lattice and the ones on each node of the layers of the 2D lattices. 
The Hilbert space of the qubit is denoted by $\ket{a}_i$ with $i=0,1,2,3$ and $a=0,1$, where the index $i$ distinguishes the qubits on the cube ($i=0$) and the ones on the 2D lattices stacked in the $x(i=1)$-, $y(i=2)$-, and $z(i=3)$-directions. 
We define the following paramagnet Hamiltonian:
\begin{eqnarray}
    H=-\sum_{\br}\left( \tau^X_{0,\br+\frac{\jx}{2}}+\tau^X_{0,\br+\frac{\jy}{2}}+\tau^X_{0,\br+\frac{\jz}{2}} \right)
    -\sum_{\br}\sum_{i=1}^3\tau^X_{i,\br+\frac{\jx}{2}+\frac{\jy}{2}+\frac{\jz}{2}}, \label{para}
\end{eqnarray}
where $\jz :=(0,0,1)$, $\tau^X_{0,\br+\frac{\jx}{2}}$ is the Pauli operator acting on the qubit at the link of the cubic lattice with the coordinate~$\br+\frac{\jx}{2}$ and the other Pauli operators are similarly defined. 
\par
We think of implementing the CZ gates~\eqref{cz} on qubits which are located at a face of the cubic lattice and the ones on the adjacent layer of the 2D lattice.
Instead of writing explicitly the terms, we make use of visual illustrations given in Figs.~\ref{xcube1} and~\ref{xcube2}.
The CZ gates act on qubits on a 2D lattice, say the one stacked along the $x$-direction, and the closest face of the cube in the $+x$-direction, parallel to the $yz$-plane. In Fig.~\ref{xcube1}, they correspond to the layer and face marked by the yellow arrow. Viewing from the $x$-axis, the layer and face form a 2D lattice where qubits are placed on each node and link, as shown in Fig.~\ref{xcube2}.
The CZ gates act on a qubit defined on a node and four qubits that surround it, as well as on a qubit located on a link and two qubits that are connected with it. 
The former corresponds to $a\sim e$ in the left of Fig.~\ref{xcube2} and the latter does to $a$, $d,g$, and $a,e,f$ in the right of Fig.~\ref{xcube2}. 
Such manipulation reminds us of the realization of cluster states of the SPT phase preserving the $0$-form and $1$-form global symmetries.
The CZ gates are implemented on qubits defined on a different layers stacked along the $y$- and $z$-directions and the ones on the closest faces in the similar manner.

The model has the symmetries generated by the following operators
\begin{eqnarray}
 &&  \prod_{e\perp S}\tau^X_{0,e},\nonumber\\
 &&\prod_{\hy,\hz}\tau^X_{1,\br+\frac{\jx}{2}+\frac{\jy}{2}+\frac{\jz}{2}}\;(1\leq \hx\leq L_x),
 \prod_{\hz,\hx}\tau^X_{2,\br+\frac{\jx}{2}+\frac{\jy}{2}+\frac{\jz}{2}}\;(1\leq \hy\leq L_y),\;
 \prod_{\hx,\hy}\tau^X_{3,\br+\frac{\jx}{2}+\frac{\jy}{2}+\frac{\jz}{2}} \; (1\leq \hz\leq L_z).\label{sym}
\end{eqnarray}
Here the first product represents the multiplication of the $\tau^X_0$ operators on links that cross a closed surface~$S$ in the dual sites, which corresponds to a global $1$-form symmetry~\cite{raussendorf2005long,gaiotto2015generalized} whereas the last three products do to the subsystem symmetries.
One can gauge these symmetries. Following the same procedure outlined in Sec.~\ref{2.2}, we gauge the last three symmetries in~\eqref{sym} on each layer of the 2D lattice.\par
Regarding gauging the $1$-form $\mathbb{Z}_2$ global symmetry, 
we follow the analogous steps for gauging ordinary $\mathbb{Z}_N$ symmetries. Since the procedure closely parallels the one in the previous cases~\cite{kapustin2017higher}, we outline how to carry out gauging succinctly. We introduce $2$-form gauge fields which reside on each face $f$ of the cube with the Pauli operator $X_{0,f}, Z_{0,f}$,
and impose the Gauss law constraint on each link $e$, $G_e\ket{\rm phys}=\ket{\rm phys}$ with
\begin{equation*}
    G_{e}=\tau^X_{0,e}\prod_{f | e\subset \partial f}X_{0,f} ,
\end{equation*} 
where the product corresponds to multiplication of four Pauli operators that belong to the link.
We further minimally couple the quadratic term on links that belong to a face,  
$\prod_{e\subset \partial f}\tau^Z_{0,e}$ to the $2$-form gauge field via $Z_{0,f}\prod_{e\subset \partial f}\tau^Z_{0,e}$ and impose the flux-less condition of the 2-form gauge fields.

\par
%
\begin{figure}[th]
    \begin{center}
        \begin{subfigure}[h]{0.36\textwidth}
  \includegraphics[width=\textwidth]{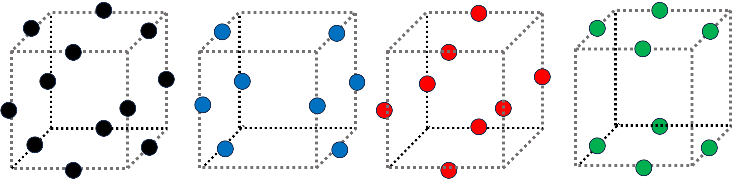}
         \caption{}\label{xcc}
             \end{subfigure}
               \begin{subfigure}[h]{0.64\textwidth}
  \includegraphics[width=\textwidth]{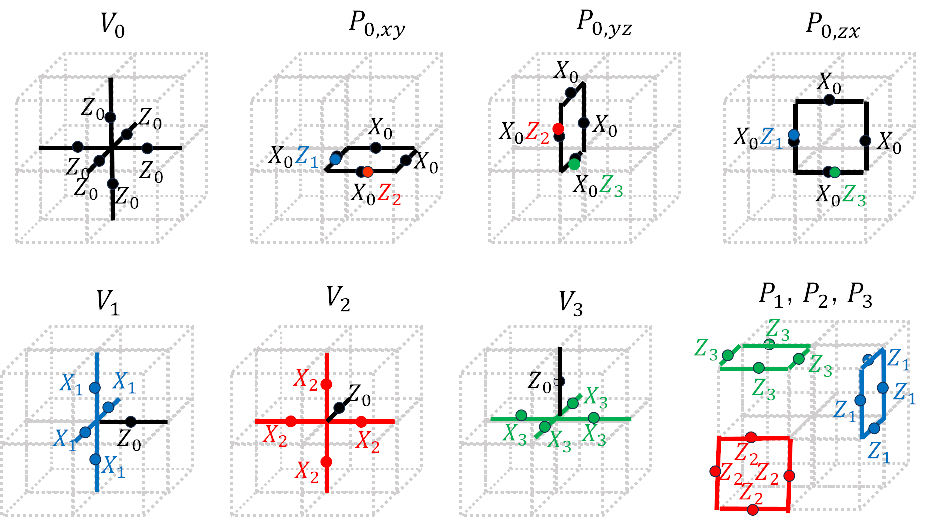}
         \caption{}\label{xcube3}
             \end{subfigure}
                  \begin{subfigure}[h]{0.60\textwidth}
  \includegraphics[width=\textwidth]{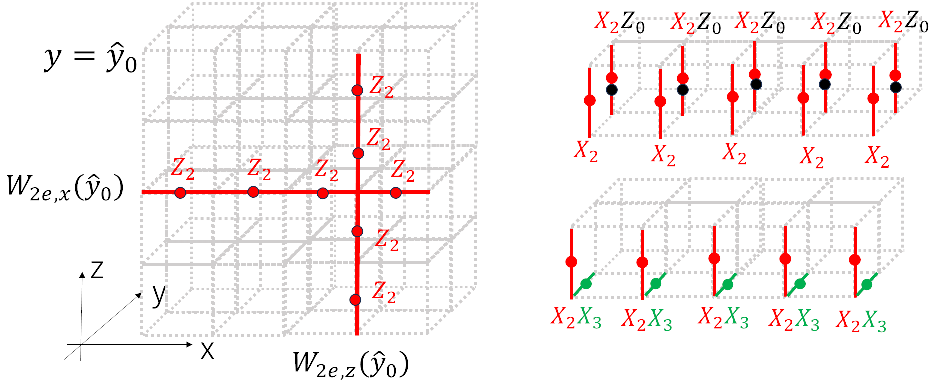}
         \caption{}\label{xc4}
             \end{subfigure}
   \end{center}
\caption{(a) Configurations of the qubits after gauging and redefining the lattice grid. (b) The terms defined in~\eqref{74} and \eqref{75} that constitute the Hamiltonian~\eqref{lattice fl}. (c) Configurations of loops given in~\eqref{loop3}.    }
   \end{figure}

After gauging the symmetries~\eqref{sym}, we also redefine the lattice grid to make the figure more visually-friendly. We do so in such a way that gauge fields are located on each link of the newly-defined cubic lattice; 
the gauge fields introduced by implementing  gauging the first symmetry in~\eqref{sym} is located on links of the cubic lattice (black dots in Fig.~\ref{xcc}) whereas the ones introduced by the second, third, and fourth symmetry in~\eqref{sym} distinguished by the index $i=1$, $i=2$ and $i=3$, is placed on links belonging to the face of the cubic lattice, in the $yz$-, $zx$-, and $xy$-planes, respectively (blue, red, and green dots in Fig.~\ref{xcc}). \par
We finally arrive at the following gauged Hamiltonian:
\begin{eqnarray}
    H_{{\rm gauged},Xc}
    =&-&\biggl[\sum_{\br}V_{0,\br}+\sum_{\hz}\sum_{\bp_{xy}(\hz)}B_{0,\bp_{xy}(\hz)}+\sum_{\hx}\sum_{\bp_{yz}(\hx)}B_{0,\bp_{yz}(\hx)}+\sum_{\hy}\sum_{\bp_{zx}(\hy)}B_{0,\bp_{zx}(\hy)}\biggr]\nonumber\\
    &-&\sum_{\hx}\biggl[\sum_{\hy,\hz}V_{1,\br}+\sum_{\bp_{yz}(\hx)}B_{1,\bp_{yz}(\hx)}\biggr]
     -\sum_{\hy}\biggl[\sum_{\hx,\hz}V_{2,\br}+\sum_{\bp_{zx}(\hy)}B_{2,\bp_{zx}(\hy)}\biggr]
     -\sum_{\hz}\biggl[\sum_{\hx,\hy}V_{3,\br}+\sum_{\bp_{xy}(\hz)}B_{3,\bp_{zx}(\hz)}\biggr]. \nonumber\\
     \label{lattice fl}
\end{eqnarray}
The individual term of the first line in~\eqref{lattice fl} is defined by
\begin{eqnarray}
    V_{0,\br} &\vcentcolon=& \prod_{t=\pm1}Z_{0,\br+t\frac{\jx}{2}}Z_{0,\br+t\frac{\jy}{2}}Z_{0,\br+t\frac{\jz}{2}},\nonumber\\
   B_{0,\bp_{xy}(\hz)} &\vcentcolon=& Z_{1,\bp_{xy}(\hz)-\frac{\jx}{2}}Z_{2,\bp_{xy}(\hz)-\frac{\jy}{2}}
   \prod_{t=\pm1}
   X_{0,\bp_{xy}(\hz)+t\frac{\jx}{2}}X_{0,\bp_{xy}(\hz)+t\frac{\jy}{2}},\nonumber\\
       B_{0,\bp_{yz}(\hx)} &\vcentcolon=& Z_{2,\bp_{yz}(\hx)-\frac{\jz}{2}}Z_{3,\bp_{yz}(\hx)-\frac{\jy}{2}}\prod_{t=\pm1}X_{0,\bp_{yz}(\hx)+t\frac{\jy}{2}}X_{0,\bp_{yz}(\hx)+t\frac{\jz}{2}},\nonumber\\
       B_{0,\bp_{zx}(\hy)} &\vcentcolon=& Z_{3,\bp_{zx}(\hy)-\frac{\jx}{2}}Z_{1,\bp_{zx}(\hy)-\frac{\jz}{2}}\prod_{t=\pm1}X_{0,\bp_{zx}(\hy)+t\frac{\jx}{2}}X_{0,\bp_{zx}(\hy)+t\frac{\jz}{2}},
       \label{74}
\end{eqnarray}
which are depicted in the first four terms in Fig.~\ref{xcube3}. 
Here, we have defined the coordinate of a
plaquette in the $xy$, $yz$ and $zx$-plane by $\bp_{xy}(\hz)=(\hx+\fr,\hy+\fr,\hz)$, $\bp_{yz}(\hx)=(\hx,\hy+\fr,\hz+\fr)$ and $\bp_{zx}(\hy)=(\hx+\fr,\hy,\hz+\fr)$.
Also, the individual term of the second, third and fourth line in~\eqref{lattice fl} reads
\begin{eqnarray}
    V_{1,\br} \vcentcolon=Z_{0,\br+\frac{\jx}{2}}\prod_{t=\pm1}X_{1,\br+t\frac{\jy}{2}}X_{1,\br+t\frac{\jz}{2}},&&\quad
    B_{1,\bp_{yz}(\hx)}\vcentcolon=\prod_{s=\pm1}Z_{1,\bp_{yz}(\hx)+s\frac{\jy}{2}}Z_{1,\bp_{yz}(\hx)+s\frac{\jz}{2}},\nonumber\\
    V_{2,\br}\vcentcolon=Z_{0,\br+\frac{\jy}{2}}\prod_{t=\pm1}X_{2,\br+t\frac{\jx}{2}}X_{2,\br+t\frac{\jz}{2}},&&\quad
    B_{2,\bp_{zx}(\hy)}\vcentcolon=\prod_{s=\pm 1}Z_{2,\bp_{zx}(\hy)+s\frac{\jx}{2}}Z_{2,\bp_{zx}(\hy)+s\frac{\jz}{2}},\nonumber\\
      V_{3,\br}\vcentcolon=Z_{0,\br\frac{\jz}{2}}\prod_{t=\pm1}X_{3,\br+t\frac{\jx}{2}}X_{3,\br+t\frac{\jz}{2}},&&\quad
      B_{3,\bp_{xy}(\hz)}\vcentcolon=\prod_{s=\pm1}Z_{3,\bp_{xy}(\hz)+s\frac{\jx}{2}}Z_{3,\bp_{xy}(\hz)+s\frac{\jy}{2}}.\label{75}
\end{eqnarray}
These terms are portrayed in the last four terms in Fig.~\ref{xcube3}.\par
The ground state of the Hamiltonian~\eqref{lattice fl} is the projected state, satisfying that all of the eigenvalues of the terms in~\eqref{74} and \eqref{75} are trivial, e.g., $ V_{0,\br}=1$.
Essentially, the first line in~\eqref{lattice fl} describes the 3D toric code whereas the terms in the second line in~\eqref{lattice fl} do the layers of the 2D toric codes stacked along the $x$-, $y$-, and $z$-directions. 
The crucial difference between the toric codes and the model~\eqref{lattice fl} is that a few Pauli operators are multiplied with the plaquette and vertex terms of the toric codes, which alter the statistics of the fractional excitations, mirroring the fact that we accommodate the foliated DW twist terms.
\par
We investigate the GSD of the model~\eqref{lattice fl} on a 3D torus with system size $L_x\times L_y\times L_z$ by counting the number of the distinct non-contractible loops of the electric charges.
Practically, we evaluate
the number of the distinct operators of the Pauli $Z_i$ operators that form the non-contractible loops.
The model admits a number of non-contractible loops constructed by the product of the Pauli $Z_i$ operators: 
\begin{eqnarray}
    W_{1e,y}(\hx) = \prod_{\hy=1}^{L_y}Z_{1,\br+\frac{\jy}{2}},&& \quad 
    W_{1e,z}(\hx)=\prod_{\hz=1}^{L_z}Z_{1,\br+\frac{\jz}{2}}\qquad (1\leq \hx\leq L_x),\nonumber\\
    W_{2e,z}(\hy) = \prod_{\hz=1}^{L_z}Z_{2,\br+\frac{\jz}{2}},&&\quad 
    W_{2e,x}(\hy)=\prod_{\hx=1}^{L_x}Z_{2,\br+\frac{\jx}{2}}\qquad  (1\leq \hy\leq L_y),\nonumber\\
    W_{3e,x}(\hz) = \prod_{\hx=1}^{L_x}Z_{3,\br+\frac{\jx}{2}},&&\quad 
    W_{3e,y}(\hz) = \prod_{\hy=1}^{L_y}Z_{1,\br+\frac{\jy}{2}}\qquad  (1\leq \hz\leq L_z).\label{88}
\end{eqnarray}
Examples are shown in the left of Fig.~\ref{xc4}. Naively there are  ${2L_x+2L_y+2L_z}$ distinct loops, however, there are a few constraints on them, reducing the number.\footnote{
One might wonder that the model admits non-contractible membrane operators which are found in the 3D toric code, such as
$ W_{0e,xy}=\prod_{\hx=1}^{L_x}\prod_{\hy=1}^{L_y}Z_{0,\br+\frac{\jz}{2}}$. However, 
such membranes become trivial as multiplication of the terms in~\eqref{75} gives identity. Indeed, we have
$     1=\prod_{\hy=1}^{L_y}\prod_{\hz=1}^{L_z}V_{1,\br}=W_{0e,xy}$.}
Indeed, multiplication of the operators in~\eqref{74} gives
\begin{eqnarray}
1 &=& \prod_{\hx,\hy}    B_{0,\bp_{xy}(\hz)}=\prod_{\hy}W_{2e,x}(\hy)\times\prod_{\hx}W_{1e,y}(\hx),\nonumber\\
1 &=& \prod_{\hy,\hz} B_{0,\bp_{yz}(\hx)}=\prod_{\hz}W_{3e,y}(\hz)\times\prod_{\hy}W_{2e,z}(\hy), \nonumber\\
1 &=& \prod_{\hz,\hx}B_{0,\bp_{zx}(\hy)}=\prod_{\hz}W_{3e,x}(\hz)\times\prod_{\hx}W_{1e,z}(\hx),\label{89}
\end{eqnarray}
where we have used the fact that the ground states satisfies that the eigenvalues of the operators~\eqref{74} are trivial. Such constraints come from the fact that a few Pauli operators are attached to the terms in the 2D toric codes, due to the foliated DW twist terms that we introduced.
Hence, the distinct number of non-contractible loops of the $\mathbb{Z}_2$ electric charges is given by ${2L_x+2L_y+2L_z}-3$, leading to
\begin{equation}
    {\rm GSD} =2^{2L_x+2L_y+2L_z-3},\label{gsdxc}
\end{equation}
which agrees with the GSD found in~\cite{Vijay}.\par
One can analogously count the number of distinct loops of the magnetic charges, i.e., the number of non-contractible loops constructed by $X_i$'s. Examples of such loops are
\begin{eqnarray}
    W_{2m,x}(\hy)=\prod_{\hx}X_{2,\br+\jy+\frac{\jz}{2}}X_{2,\br+\frac{\jz}{2}}Z_{0,\br+\jy+\frac{\jz}{2}},\quad
    W_{2m,3m,x}(\hy)=\prod_{\hx}X_{2,\br+\frac{\jz}{2}}X_{3,\br+\frac{\jy}{2}}\label{loop3}
\end{eqnarray}
which are depicted in Fig.~\ref{xc4}. We can find other loops in the similar form.
By the similar argument around~\eqref{89}, it can be verified that 
there are $2L_x+2L_y+2L_z-3$ distinct loops. By using these loops as well as the ones we have considered~\eqref{88}, one can make $2L_x+2L_y+2L_z-3$ pairs of loops each of which generates the $\mathbb{Z}_2$ Heisenberg algebra, arriving at the GSD~\eqref{gsdxc}.

\section{Discussion and conclusion}
\label{s6}
Originally proposed in the context of quantum information~\cite{chamon,Haah2011}, the fracton topological phases have now diverse research interests, involving various branches of physics, such as condensed matter physics, and high-energy physics. Motivated by the ultimate goal to construct a complete theoretical framework to describe such phases, the concept of symmetries has been developed. 
Focusing on one of new symmetries, multipole symmetry, we have demonstrated a way to build up topological phases which respect such a symmetry. Take-home message in this work is that via gauging SPT phases with global $\mathbb{Z}_N$ symmetries, they become the foliated DW twist terms which yield unconventional fractional excitations found in the multipole topological phases. Furthermore, this construction 
gives a clear physical understanding of the term $A^I\wedge b\wedge e^I$ in a field theory description of the multipole topological orders. The way we construct phases can also be implemented in other fracton models. Before closing this section, we make a few remarks and future research directions, regarding the present work.

\subsection{Comments on entanglement entropy}
Recently, several works studied entanglement entropy of topological phases with multipole symmetries. It is widely known that the entanglement entropy of a contractible disk geometry $A$ in a conventional $(2+1)$d topologically ordered phase reads $S_A=A_{\rm area}-\gamma$~\cite{LevinWen2006TEE,KitaevPreskill2006TEE}. While the first term is the so-called area law term, proportional to the perimeter of the disk, the second sub-leading order term is particularly important. It is known as the topological entanglement entropy which relates to a topological order via $\gamma=\log\sqrt{\sum_ad_a^2}$, where $d_a$ denotes a quantum dimension of an anyon labeled by $a$. 
It was shown that the topological entanglement entropy of a contractible disk geometry in a $\mathbb{Z}_N$ topological phase with multipole symmetries gives $\gamma=m\log N$ ($m\in\mathbb{Z}$), which is in contrast with the fact that the number of superselection sectors, that is, the number of distinct fractional excitations depends on the greatest common divisor between $N$ and the system size~\cite{ebisu2302ee,2023arXiv231009425K}. \par
Such a result can be understood by the form of the BF theory of topological phases with multipole symmetries~\eqref{foliation dipole} and \eqref{foliation dipole_xy} together with the interpretation of the coupling term that we make throughout this paper, i.e., interpretation of the term $A^I\wedge b\wedge e^I$ as the foliated DW twist terms. 
Generally, it was shown that the DW twist terms alter the fractional statistics of anyons but do not change the total quantum dimension. 
Indeed, it was proven that in a concrete lattice model of a quantum double with DW twist terms the topological entanglement entropy of a disk geometry takes the same values as the one in the case without the DW twist~\cite{PhysRevB.97.085147}. 
By this result, together with our observation that the BF theory of the topological phases with multipole symmetries consists of layers of the toric codes with the foliated DW twist terms, leads to that the topological entanglement entropy of such phases has the form $\gamma=k\log N$, where $k$ is the number of layers of the toric codes (more explicitly, the number of degrees of freedom of the multipoles such as global charge, dipole, quadrupole, e.t.c.).\par
As an example, the topological entanglement entropy of a disk in the UV stabilizer models respecting the dipole symmetry in the $x$-direction was found to be $\gamma=2\log N$~\cite{ebisu2302ee}, which is consistent with the form of the BF theory~\eqref{foliation dipole} as the theory consists of the two copies of the toric codes with the DW twist; regardless of the presence of the DW twist, the total quantum dimension of the excitations amounts to be the one of the two copies of the toric codes, hence $\gamma=2\log N$.

\subsection{Future directions}
There are several future directions regarding the present work.
Studying topological phases with multipole symmetries in view of quantum information, such as quantum error correction~\cite{dennis2002topological}, is a widely open problem. Especially, it could be theoretically and practically important to address how stable the model considered in this work is against errors e.g., bit or phase flips and how the multipole symmetry plays a role compared with the conventional topological stabilizer codes. Investigating observable in the context of quantum information other than entanglement entropy, such as quantum negativity~\cite{Vidal_negativity}, could be an another interesting direction.
\par
It would be interesting to extend our analysis to the case of arrays of SPT phases
with general group symmetry $G$ to realize more general foliated DW twist terms, as our approach nicely fits into the group cohomology. Especially, in the case of $G$ being non-Abelian, the investigation on the compatibility condition with the periodic boundary condition becomes complicated, which could yield richer multipole or fracton topological phases. \par
It would also be intriguing to incorporate other types of symmetries into our construction. The gauge transformation of the multipole symmetries~\eqref{gaugetr1} resembles the one found in $2$-group symmetry~\cite{Cordova:2018cvg} in that different gauge fields are related with one another via gauge transformation. Studying whether one can construct topological phases with higher group symmetry, or even richer topological phases where both of higher-group and multipole symmetries are incorporated by our approach would contribute to making better understanding of new topological phases with various symmetries. 

Recently it was found that several field theories related to fractons have 't Hooft anomalies and their anomaly inflows \cite{Burnell:2021reh,Yamaguchi:2021xeq,Honda:2022shd,Cao:2023doz,Huang:2023zhp,Cao:2023rrb}.
It would be illuminating to study relations among them and foliated BF theories.
Last but not least, some foliated BF theories themselves may also have 't Hooft anomalies. 
It would be also interesting to study this aspect in future.

\section*{Acknowledgement}
We thank Qiang Jia, Ken Shiozaki and Shimamori Souichiro for helpful discussion.
H.~E. is supported by KAKENHI-PROJECT-23H01097.
M.~H. is supported by MEXT Q-LEAP, JSPS Grant-in-Aid 
for Transformative Research Areas (A) ``Extreme Universe" JP21H05190 [D01] and JSPS KAKENHI Grant Number 22H01222.
M.~H. and T.~N. are supported by JST PRESTO Grant Number JPMJPR2117.
T.~N is supported by JST, the establishment of university fellowships towards the creation of science technology innovation, Grant Number
JPMJFS2123.

\bibliography{main}

\end{document}